\documentclass[a4paper,fleqn,usenatbib]{mnras}

\usepackage{newtxtext,newtxmath}
\usepackage[T1]{fontenc}
\usepackage{ae,aecompl}

\usepackage{graphicx}	% Including figure files
\usepackage{amsmath}	% Advanced maths commands
\usepackage{natbib}    %bibtex
\usepackage{threeparttable}
\usepackage{chngpage}
\usepackage{color}
\usepackage{xcolor}
\usepackage{ulem}

\newcommand{\Msun}{M_{\odot}}

\newcommand{\Fefs}{$^{56}$Fe}
\newcommand{\Cofs}{$^{56}$Co}
\newcommand{\Nifs}{$^{56}$Ni}

\title[SN 2021gmj]{
Intermediate-luminosity Type IIP SN 2021gmj: a low-energy explosion with signatures of circumstellar material \\
}

\author[Y. Murai et al.]
{Yuta Murai,$^{1}$\thanks{E-mail: yuta.murai@astr.tohoku.ac.jp}
Masaomi Tanaka,$^{1,2}$ 
Miho Kawabata,$^{3}$ 
Kenta Taguchi,$^{4}$ 
Rishabh Singh Teja,$^{5,6}$ 
\newauthor
Tatsuya Nakaoka,$^{7}$ 
Keiichi Maeda,$^{4}$ 
Koji S. Kawabata,$^{7,8}$ 
Takashi Nagao,$^{9,10,11}$ 
\newauthor
Takashi J. Moriya,$^{12,13}$ 
D. K. Sahu,$^{5}$ 
G. C. Anupama,$^{5}$ 
Nozomu Tominaga,$^{12,14,15}$ 
\newauthor
Tomoki Morokuma,$^{16,17,18}$ 
Ryo Imazawa,$^{8,7}$ 
Satoko Inutsuka,$^{4}$ 
Keisuke Isogai,$^{19,20}$ 
\newauthor
Toshihiro Kasuga,$^{12}$ 
Naoto Kobayashi,$^{21}$ 
Sohei Kondo,$^{21}$ 
Hiroyuki Maehara,$^{22}$ 
Yuki Mori,$^{21}$ 
\newauthor
Yuu Niino,$^{21}$
Mao Ogawa,$^{4}$ 
Ryou Ohsawa,$^{12}$ 
Shin-ichiro Okumura,$^{23}$ 
Sei Saito,$^{1}$ 
\newauthor
Shigeyuki Sako,$^{18,24,25}$ 
Hidenori Takahashi,$^{21}$ 
Kohki Uno,$^{4}$ 
Masayuki Yamanaka$^{26}$ 
\vspace{0.2cm} \\ 
Affiliations are at the end of the paper.
}
\begin{document}
\label{firstpage}
\pagerange{\pageref{firstpage}--\pageref{lastpage}}
\maketitle

% Abstract of the paper
\begin{abstract}
We present photometric, spectroscopic and polarimetric observations of the intermediate-luminosity Type IIP supernova (SN) 2021gmj from 1 to 386 days after the explosion. 
The peak absolute $V$-band magnitude of SN 2021gmj is $-15.5$ mag, 
which is fainter than that of normal Type IIP SNe. The spectral evolution of SN 2021gmj resembles that of other sub-luminous supernovae: the optical spectra show narrow P-Cygni profiles, indicating a low expansion velocity. We estimate the progenitor mass to be about 12 $\rm \Msun$ from the nebular spectrum and the \Nifs~mass to be about 0.02 $\rm \Msun$ from the bolometric light curve. We also derive the explosion energy to be about 3$\times10^{50}$ erg by comparing numerical light curve models with the observed light curves. 
Polarization in the plateau phase is not very large, suggesting nearly spherical outer envelope. 
The early photometric observations capture the rapid rise of the light curve, which is likely due to the interaction with a circumstellar material (CSM). 
The broad emission feature formed by highly-ionized lines on top of a blue continuum in the earliest spectrum gives further indication of the CSM at the vicinity of the progenitor. 
Our work suggests that a relatively low-mass progenitor of an intermediate-luminosity Type IIP SN can also experience an enhanced mass loss just before the explosion, as suggested for normal Type IIP SNe.
\end{abstract}

% Select between one and six entries from the list of approved keywords.
% Don't make up new ones.
\begin{keywords}
supernovae: individual (SN 2021gmj)
\end{keywords}

%%%%%%%%%%%%%%%%%%%%%%%%%%%%%%%%%%%%%%%%%%%%%%%%%%

%%%%%%%%%%%%%%%%% BODY OF PAPER %%%%%%%%%%%%%%%%%%

\section{Introduction}
\label{sec:intro}

Core-collapse supernovae (CCSNe) are the explosions of massive stars 
with a zero-age main-sequence (ZAMS) mass of $\gtrsim 8 ~ \rm \Msun$ 
(e.g., \citealt{Heger2003}). 
CCSNe with hydrogen lines in their spectra around the maxima 
of the brightness are classified as Type II SNe 
(\citealt{Woosley1986, Filippenko1997}). 
Among them, SNe that display a plateau feature (\citealt{Barbon1979}) 
in their light curves are classified as Type IIP SNe, 
which are the most common SNe among CCSNe 
($\sim60\%$ of CCSNe; \citealt{Li2011}). 

Many studies have revealed that Type IIP SNe have a wide range 
of plateau luminosities ($-18\lesssim M_{V \rm , plateau}\lesssim-14 ~ \rm mag$, 
where $M_{V \rm , plateau}$ is the absolute magnitude during the plateau phase in the $V$ band;
\citealt{Anderson2014, Valenti2016}). 
The observational diversity of Type IIP SNe has been revealed 
not only in the luminosity but also in the synthesized \Nifs ~mass 
($0.001 \lesssim M_{\rm Ni} \lesssim 0.360 ~ \rm \Msun$; \citealt{Muller2017, Anderson2019, Rodriguez2021, Martinez2022}) 
and the ejecta velocities at 50 days post-explosion
($1,500 \lesssim v_{\rm ej} \lesssim 9,600 ~ \rm km ~ s^{-1}$; 
\citealt{Gutierrez2017}). 
In particular, sub-luminous, low-velocity and \Nifs-poor 
Type IIP SNe such as SN 2005cs (\citealt{Pastorello2006}) 
have been of interest to understand the variety of the population of Type IIP SNe toward the faintest end. 
In fact, sub-luminous Type IIP SNe are considered to be lower-energy explosions 
of lower-mass progenitors as compared to normal Type IIP SNe 
(\citealt{Maund2005, Tsvetkov2006, Pastorello2009, Smartt2009}). 
Moreover, some studies suggest that such sub-luminous SNe can be explained by 
electron-capture supernovae (\citealt{Hosseinzadeh2018, Hiramatsu2021, Valerin2022}). 
In addition, Type IIP SNe that have properties 
lying between those of normal and sub-luminous objects have also been discovered 
(e.g., SN 2008in, \citealt{Roy2011}; SN 2009ib, \citealt{Takats2015}; SN 2009N, 
\citealt{Takats2014}; SN 2012A, \citealt{Tomasella2013}). 

Recent high-cadence transient surveys enable us to probe 
the final evolutionary stage of SN progenitors. 
Rapid photometric observations suggest that 
a large fraction of massive stars undergo an enhanced mass loss 
producing extensive circumstellar material (CSM) surrounding these stars 
in the final years before explosion 
(e.g., \citealt{Valenti2016, Morozova2017, Morozova2018, Forster2018}). 
Early spectra observed within a few days after explosion also imply 
the existence of extensive CSM around the progenitors, 
showing highly-ionized narrow emission lines 
produced by photoionization due to the intensive ultraviolet radiation 
from SNe (\citealt{Gal-Yam2014, Smith2015, Khazov2016, Yaron2017}). 
Such early spectra have also been studied for sub-luminous SNe. 
For instance, \cite{Hosseinzadeh2018} and \cite{nakaoka2018} reported a feature around 4600 {\AA} 
in the earliest spectrum of a sub-luminous Type IIP SN 2016bkv, 
attributing it to He II, C III, and N III originating from CSM. 
Similarly, the sub-luminous Type IIP SN 2018lab shows 
such highly-ionized emission lines due to CSM 
in its early spectra (\citealt{Pearson2023}). 
However, the sample of such early observations 
for intermediate-luminosity or sub-luminous Type IIP SNe are still limited 
as the observational sample is smaller than normal Type IIP SNe.

The mechanism of the enhanced mass-loss forming such dense CSM 
has also been under debates. 
For instance, \cite{Yoon2010} proposed pulsations of red supergiants (RSGs) drive superwinds and result in the enhanced mass loss. 
\cite{Moriya2014} proposed core neutrino emission shortly before explosion 
makes luminosity of stars Eddington luminosity, resulting in the enhanced mass loss. 
\cite{Quataert2016} proposed super-Eddington stellar winds driven 
by near-surface energy deposition causes the enhanced mass loss. 
In order to understand the mechanism of the enhanced mass loss, 
it is important to study the CSM for a variety of supernova progenitors. 
In fact, some scenarios invoke nuclear flash near the end of stellar evolution, 
which is expected to be more important for lower-mass progenitors \citep{Woosley2015}. 
Thus, it is of interest to study the CSM around lower-luminosity SNe likely with 
lower-mass progenitors. 

In this paper, we report our photometric, spectroscopic and polarimetric observations 
for the intermediate-luminosity Type IIP SN 2021gmj 
from a few days to more than a year after the explosion. 
In Section \ref{sec:observation}, we present our observations and data reduction. 
In Section \ref{sec:results}, we show photometric, spectroscopic 
and polarimetric properties of SN 2021gmj 
and compare them with those of other Type IIP SNe. 
In Section \ref{sec:discussion}, we present a detailed analysis with estimates of the \Nifs ~mass, 
progenitor mass, mass-loss rate, and explosion energy of SN 2021gmj. 
Finally, we summarize our findings in Section \ref{sec:conclusions}.

\section{Observation and data reduction}
\label{sec:observation}

SN 2021gmj was discovered by the Distance Less Than 40 Mpc Survey 
(DLT40; \citealt{Tartaglia2018}) at 07:40:48 UTC on 2021-03-20 
(59293.32 MJD; \citealt{Valenti2021}). 
SN 2021gmj is located at ($\alpha$, $\delta$) = 
($10^{\rm h}38^{\rm m}47.345^{\rm s}$, $+53^{\circ}30'31.79\arcsec$), 
with an offset of 19.73$\arcsec$ from the center of its host galaxy, NGC 3310 at a redshift $z = 0.003312 \pm 0.000009$. 
The distance to the host galaxy is somewhat uncertain: 
there are some redshift-independent measurements reported in the NASA/IPAC NED Database, 
but they are primarily from the Tully-Fisher relation \citep{Bottinelli1984,bottinelli1986}. 
The weighted average distance for these measurements is 19.7 Mpc 
(distance modulus $\mu = 31.47$ mag, \citealt{Steer2021}), 
but there is a large dispersion from 17.6 Mpc to 20.2 Mpc. 
Thus, instead of using these distance estimates, 
we simply convert the redshift to the luminosity distance of $19.2$ Mpc ($\mu = 31.42$ mag), 
which is obtained in the NED by considering the influence of the Virgo cluster, the Great Attractor (GA), 
and the Shapley supercluster 
by assuming $H_{0}$ = 73.0 km $\rm s^{\rm -1} ~ Mpc^{\rm -1}$ (\citealt{Riess2022}), 
$\rm \Omega_{m}$ = 0.27 and $\rm \Omega_{\Lambda}$ = 0.73.

Throughout this paper, we show photometry corrected only for 
the Milky Way extinction ($A_{V}^{\rm MW}$ = 0.0589 mag) 
assuming $R_{V}$ = 3.1 (\citealt{Fitzpatrick1999}), 
based on the Galactic extinction map of \cite{Schlafly2011}.
We do not take the host-galaxy extinction into account as our spectra of SN 2021gmj do not show significant interstellar Na I D $\lambda$5889,5895 absorption lines,
with an upper limit of the equivalent width of 0.7496 {\AA}. 
This corresponds to $E(B-V) \lesssim 0.1$ mag (or $A_{\rm V} \lesssim 0.3$ mag) 
by using the empirical relation by \cite{Poznanski2012}.

\subsection{Photometry}
\label{subsec:photometry}

Our optical and near-infrared photometric observations of SN 2021gmj 
were performed using the following instruments and telescopes. 
The results of photometry are provided in Tables \ref{tab:opt-photometry} 
and \ref{tab:nir-photometry}.

\subsubsection{Tomo-e Gozen}
The Tomo-e Gozen camera (\citealt{Sako2018}) is mounted on 
the Kiso Schmidt telescope, which is a 1.05-m telescope at Kiso Observatory, 
the University of Tokyo in Japan. 
This camera takes contiguous images at a rate of 2 frames per second 
with a wide field-of-view (20.8 deg$^{2}$) covered by 84 CMOS sensors. 
The camera has no filter and the CMOS sensors are sensitive from 300 to 1000 nm 
with a peak efficiency of 0.72 at 500 nm. 
The images of SN 2021gmj were obtained from 2.16 days to 73.15 days 
after the discovery. 

The basic data reduction for the photometric data was performed 
using Tomo-e Gozen pipeline (\citealt{Sako2018}). 
In this process, bias, dark and flat corrections were performed. 
Then, the World Coordinate System (WCS) was assigned with \texttt{astrometry.net} (\citealt{Lang2010}). 
Zero-point magnitudes were determined using the Pan-STARRS catalog (\citealt{Chambers2016}). 
In order to properly measure the brightness of the SN, 
it is necessary to remove the background host galaxy light 
by subtracting a template image from science images. 
In the Tomo-e Gozen pipeline, Pan-STARRS $r$-band images were used as the template images. 
However, the image subtraction sometimes gives a residual 
as the wavelength response of the Pan-STARRS $r$-band images 
is not identical to that of Tomo-e Gozen images. 
Thus, in this work, we created a template image by stacking Tomo-e Gozen images 
that were obtained before the date of the last non-detection (19th March 2021), from 15th June 2020 to 18th March 2021. 
Then we used the stacked image for image subtraction 
which was performed using \texttt{HOTPANTS} (\citealt{Becker2015}). 
Finally, aperture photometry with an aperture size 
that is twice as large as the full width at half maximum (FWHM) of the object 
was performed on the subtracted images.

\subsubsection{HONIR / HOWPol}
We obtained $V$-, $R$-, $I$-, $J$-, $H$- and $Ks$-band photometry of SN 2021gmj using the Hiroshima Optical and Near-Infrared camera (HONIR; \citealt{Akitaya2014}) mounted on the Kanata telescope, 
which is a 1.5-m telescope at the Higashi-Hiroshima Observatory, Hiroshima University in Japan.
The optical and near-infrared images were obtained from 1.43 days to 130.17 days after the discovery. 

In the data reduction process, bias, dark and flat corrections were performed using standard procedures in \texttt{IRAF} (\citealt{Tody1986}). 
For determining the zero points in these images, 
the magnitudes of comparison stars taken from the Pan-STARRS catalog 
were used by converting the $g$-, $r$- and $i$-band magnitudes 
into the $V$-, $R$- and $I$-band ones using the relation in \cite{Jordi2006}. 
For the NIR comparison stars, magnitudes were taken from the 2MASS catalog (\citealt{Persson1998}). 
In order to subtract a template image from the object images, 
we used HONIR images that were obtained at the last observational epoch as the template images, 
in which the SN was not detected.
We measured the limiting magnitudes of these images obtained at the last epoch 
by measuring the fluxes at random sky positions (aperture radius is FWHM $\times$ 2). 
We confirmed that the limiting magnitude of these images is shallower than 
the SN brightness at the same epoch that was obtained by TriCCS (see Section \ref{subsubsec:TriCCS}).
We therefore used these images obtained at the last epoch as the template images for the HONIR images. 
Then, image subtraction was performed using \texttt{HOTPANTS} (\citealt{Becker2015}). 
Finally, aperture photometry was performed on the subtracted images 
in the same way as the Tomo-e Gozen data. 

We also obtained $B$-, $V$-, $R$- and $I$-band photometry of SN 2021gmj using the Hiroshima One-shot Wide-field Polarimeter (HOWPol; \citealt{Kawabata2008}) mounted on the Kanata telescope. 
The HOWPol images were obtained 
from 1.25 to 138.15 days after the discovery. 
We performed the point-spread function (PSF) fitting photometry
without conducting template subtraction, due to lack of appropriate template images for the HOWPol data. 
We did not use the $B$-band data after the plateau phase 
because by the comparison with HFOSC data (see Section \ref{subsubsec:HFOSC}), it turned out that they suffered from a contamination of the background emission. 
Since the contamination also caused a slight offset in the magnitudes in other bands, we show HOWPol photometry only for the overall comparison (not for the detailed comparison of color evolution, see Section \ref{subsec:photoevolution}).

\subsubsection{TriCCS}
\label{subsubsec:TriCCS}
We obtained $g$-, $r$- and $i$-band images of SN 2021gmj using the TriColor CMOS Camera and Spectrograph 
(TriCCS)\footnote{\url{http://www.o.kwasan.kyoto-u.ac.jp/inst/triccs/}} mounted on the Seimei telescope (\citealt{Kurita2020}), 
which is a 3.8-m telescope at the Okayama Observatory, Kyoto University in Japan, 
from 193.5 days to 351.18 days after the discovery. 
The standard image reduction including bias, dark and flat corrections was performed. 
As the template images for image subtraction, 
we used Pan-STARRS $g$-, $r$- and $i$-band images. 
Finally, aperture photometry was performed on the subtracted images in the same way 
as the Tomo-e Gozen data.

\subsubsection{HFOSC}
\label{subsubsec:HFOSC}
We obtained Bessel $U$-, $B$-, $V$-, $R$- and $I$-band images of SN~2021gmj using the Himalayan Faint Object Spectrograph Camera (HFOSC; \citealt{Prabhu2014}) mounted on the 2-m Himalayan Chandra Telescope (HCT) telescope at the Indian Astronomical Observatory (IAO), Hanle, India, from 2.18 days to 245.18 days after the discovery.

For data processing, object frames were bias subtracted, flat fielded, cosmic ray corrected, 
and WCS aligned utilizing standard tasks in \texttt{IRAF} \citep{Tody1986} along with other packages, 
namely \texttt{Astroscrappy} \citep{ vanDokkum2001, astroscrappy}, \texttt{astrometry.net} \citep{Lang2010} and \texttt{SWarp} \citep{Bertin2010}. 
Nightly zero points for individual frames were estimated by calibrating secondary standards 
in the SN field from the four standard fields \citep[PG0918+029, PG0942-029, PG1047+003, and PG1323-085:][]{Landolt1992} 
observed over two photometric nights. 
We performed aperture photometry over template-subtracted images. 
The template images were taken on 24th Feb 2023, when the SN faded significantly. 
The photometry was performed using a PyRAF-based pipeline \citep[\texttt{RedPipe,}][]{Singh2021} 
with template subtraction steps described in \citet{Singh2019}.

\subsection{Spectroscopy}
\label{subsec:spectroscopy}

We obtained optical spectra of SN 2021gmj using the following instruments and telescopes. 
The logs of spectroscopic observations are provided in Table \ref{tab:spectra}.

\subsubsection{KOOLS-IFU}

The Kyoto Okayama Optical Low-dispersion Spectrograph with optical-fiber Integral Field Unit 
(KOOLS-IFU; \citealt{Matsubayashi2019}) is mounted on the Seimei telescope. 
We used VPH-blue grism, giving a wavelength coverage of 4100-8900 {\AA} and a 
spectral resolution of $R = \lambda / \Delta\lambda \sim 500$. 
General data reduction for our spectroscopic data frames was performed with 
the procedures\footnote{\url{http://www.kusastro.kyoto-u.ac.jp/~iwamuro/KOOLS/}} 
that are developed for KOOLS-IFU data. 
We used the Image Reduction and Analysis Facility (IRAF; \citealt{Tody1986}; \citealt{Tody1993}) 
through the Python package (PyRAF; \citealt{STScI2012}) in the data reduction. 
The bias estimated from the overscan regions was subtracted 
from all of the 2D object frames of the spectral data. 
After bias subtraction, each object frame was corrected for distortion. 
Then, object frames were corrected with a flat field frame. 
We took dome-flat and twilight-flat images at each epoch 
and merged them into a flat field frame. 
For the wavelength calibration, we used Hg, Ne, and Xe lamp frames. 
Then, cosmic ray events in all of the 2D object frames were removed. 
Finally, extraction of a one-dimensional spectrum from the two-dimensional images 
and sky subtraction were performed. 
For the flux calibration, we used HR3454, HR5501, HR5191 and HR7596 
as spectrophotometric standard stars\footnote{\url{https://www.eso.org/sci/observing/tools/standards/spectra/stanlis.html}}.

\subsubsection{ALFOSC}

The Alhambra Faint Object Spectrograph and Camera 
(ALFOSC)\footnote{\url{http://www.not.iac.es/instruments/alfosc/}} is mounted on 
the 2.56-m Nordic Optical Telescope (NOT\footnote{\url{http://www.not.iac.es}}) 
at the Roque de los Muchachos Observatory. 
We used Grism 4, giving a wavelength coverage of 3200-9600 {\AA} and 
a spectral resolution of $\sim$ 360. 
The spectrum was reduced with 
the {\texttt alfoscgui}\footnote{FOSCGUI is a graphical user interface 
aimed at extracting SN spectroscopy and photometry obtained with FOSC-like instruments. 
It was developed by E. Cappellaro. A package description can be found 
at\url{sngroup.oapd.inaf.it/foscgui.html}} pipeline, 
which uses standard IRAF tasks to perform overscan, bias and flat-field corrections, 
as well as removing cosmic ray artefacts using {\texttt lacosmic} \citep{vanDokkum2001}. 
One-dimensional spectra were extracted using the {\texttt apall} task, and 
wavelength calibration was performed by comparison with arc lamps. 
The spectra were flux-calibrated against a sensitivity function derived from a standard star 
observed on the same night.

\subsubsection{HFOSC}
The low-resolution ($R\sim800$-1200) spectra using HFOSC were obtained with two grisms, Gr7 and  Gr8, and 1.92$\arcsec$ slit setup covering 3500-9000 \AA. 
The wavelength and flux calibration were performed with Arc lamps and spectrophotometric standards spectra, respectively,  taken during observations. The spectra were corrected for absolute flux using corresponding $U$-, $B$-, $V$-, $R$- and $I$-band photometry. Standard IRAF tasks were utilized for the data reduction with step-wise details mentioned in \citet{Teja_2023}.

\subsection{Imaging polarimetry}

We obtained the $V,R$-band imaging polarimetry of SN 2021gmj 
ranging from 13.7 days to 111.6 days after the discovery with ALFOSC. 
The observing log is shown in Table \ref{tab:polarimetry}. 
We obtained the linear polarimetry of the SN 
using a half-wave retarder plate (HWP) and a calcite plate. 
We adopted 4 HWP angles ($0^{\circ}$, $22.5^{\circ}$, $45^{\circ}$ and $67.5^{\circ}$).
The data were reduced and analyzed by the standard methods as described, 
e.g. in \citet[][]{Patat2006} with IRAF. 
All frames were bias subtracted and flat-field corrected. 
We performed aperture photometry on each detected source with an aperture size 
that is twice as large as the FWHM of the ordinary beam's point-spread function. 
Since the ordinary and extraordinary beams are overlapped in the images, 
the non-uniform structures of the host galaxy can create an artificial polarization signal. 
In order to assess such an error, we conducted aperture photometry using two different sky regions: 
the regions from the edge of the aperture radius to the radii 
that are three and four times as large as the FWHM. 
Then, we took the average and deviation of these measurements 
as the polarization signal and the error (in addition to the photon shot noise), respectively. 
Based on these values, we calculated the Stokes parameters, the linear polarization degree and 
the polarization angle.
When calculating the polarization degrees, 
we subtracted the polarization bias using the standard method in \citet{Wang1997}.

%%%%%%%%%%%%%%%%%%%%%%%%%%%%%%%%%%%%%%%%%%%%%%%%%%%% 
% Section: Results
%%%%%%%%%%%%%%%%%%%%%%%%%%%%%%%%%%%%%%%%%%%%%%%%%%%% 

\section{Results}
\label{sec:results}

\subsection{Photometric evolution}
\label{subsec:photoevolution}

%%%%%%%%%%%%%%%%%%%%%%%%%%%%%%%%%%%%%%%%%%%%%%%%%%%% 
% Figure: Light curve
%%%%%%%%%%%%%%%%%%%%%%%%%%%%%%%%%%%%%%%%%%%%%%%%%%%% 
\begin{figure*}
  \begin{center}
    \includegraphics[width=\textwidth]{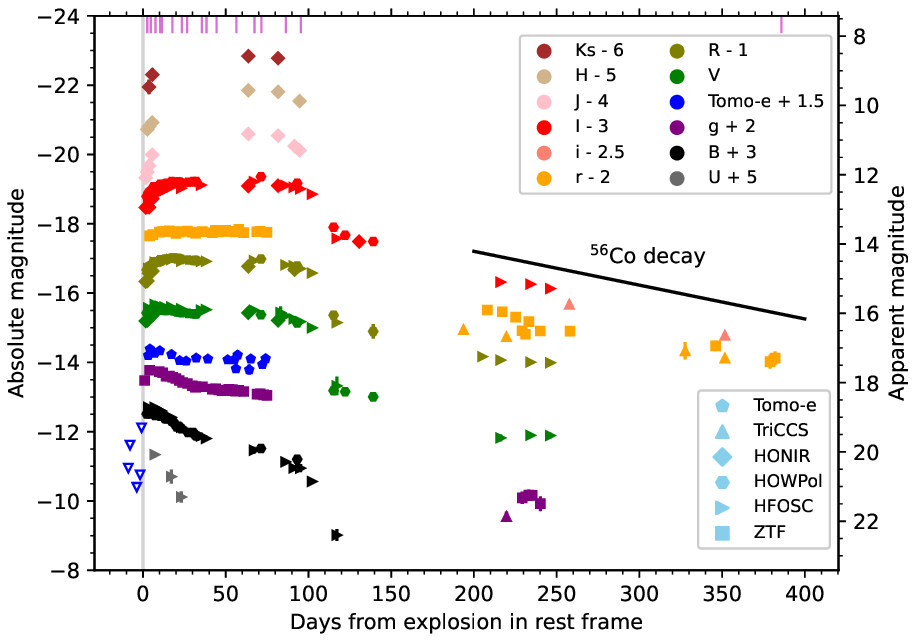}
\caption{
  \label{fig:magnitude}
  Light curves of SN 2021gmj in the optical ($U,B,g,$Tomo-e$,V,R,r,i,I$) and NIR ($J,H,K_{s}$) bands. 
  The inverse open triangles represent the upper limits. 
  The gray vertical line shows the inferred explosion date (see Section \ref{subsec:properties}). 
  The epochs of the spectroscopic observations are indicated with magenta lines on the upper x-axis. 
  The black solid line shows the slope of the \Cofs ~decay.
}
\end{center}
\end{figure*}
%%%%%%%%%%%%%%%%%%%%%%%%%%%%%%%%%%%%%%%%%%%%%%%%%%%%
%%%%%%%%%%%%%%%%%%%%%%%%%%%%%%%%%%%%%%%%%%%%%%%%%%%% 
% Figure: Compare light curve
%%%%%%%%%%%%%%%%%%%%%%%%%%%%%%%%%%%%%%%%%%%%%%%%%%%% 
\begin{figure}
  \begin{center}
    \includegraphics[width=0.47\textwidth]{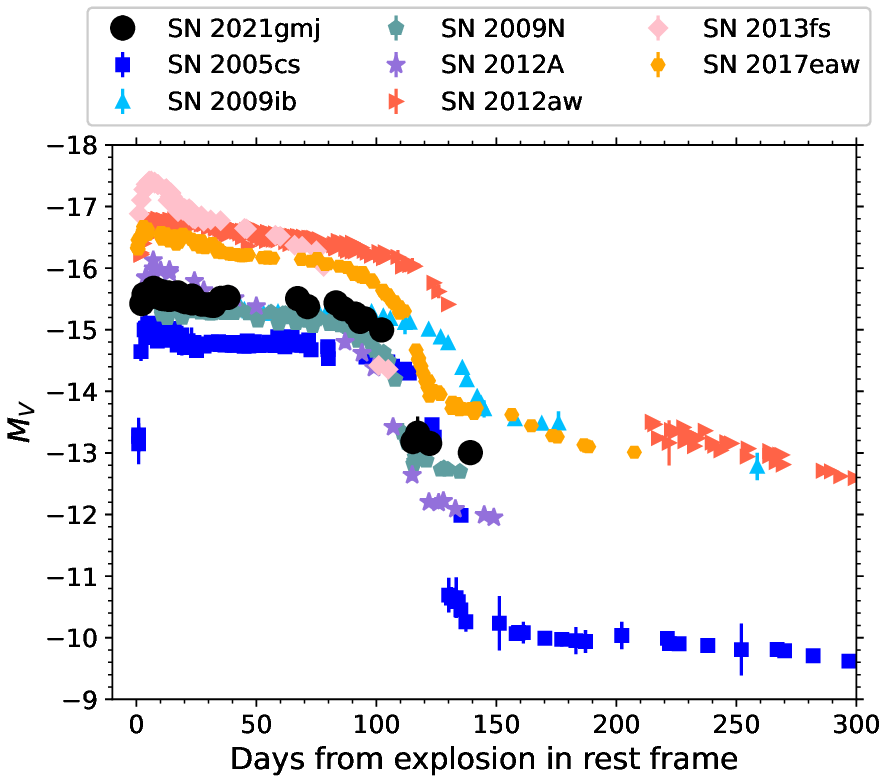}
\caption{
  \label{fig:compareMag}
  V-band light curves of SN 2021gmj and other Type IIP SNe: 
  SN 2005cs (\citealt{Tsvetkov2006, Pastorello2006, Pastorello2009}), 
  SN 2009ib (\citealt{Takats2015}), SN 2009N (\citealt{Takats2014}), 
  SN 2012A (\citealt{Tomasella2013}), SN 2012aw (\citealt{Fraser2012}), 
  SN 2013fs (\citealt{Valenti2016, Bullivant2018}) and 
  SN 2017eaw (\citealt{Tsvetkov2018}). 
  We obtained the data of the other SNe from the Open Supernova Catalog 
  (\url{https://github.com/astrocatalogs/supernovae}; \protect\citealt{Guillochon2017}). 
}
\end{center}
\end{figure}
%%%%%%%%%%%%%%%%%%%%%%%%%%%%%%%%%%%%%%%%%%%%%%%%%%%%

%%%%%%%%%%%%%%%%%%%%%%%%%%%%%%%%%%%%%%%%%%%%%%%%%%%% 
% Figure: Early compare light curve
%%%%%%%%%%%%%%%%%%%%%%%%%%%%%%%%%%%%%%%%%%%%%%%%%%%% 
\begin{figure}
  \begin{center}
    \includegraphics[width=0.47\textwidth]{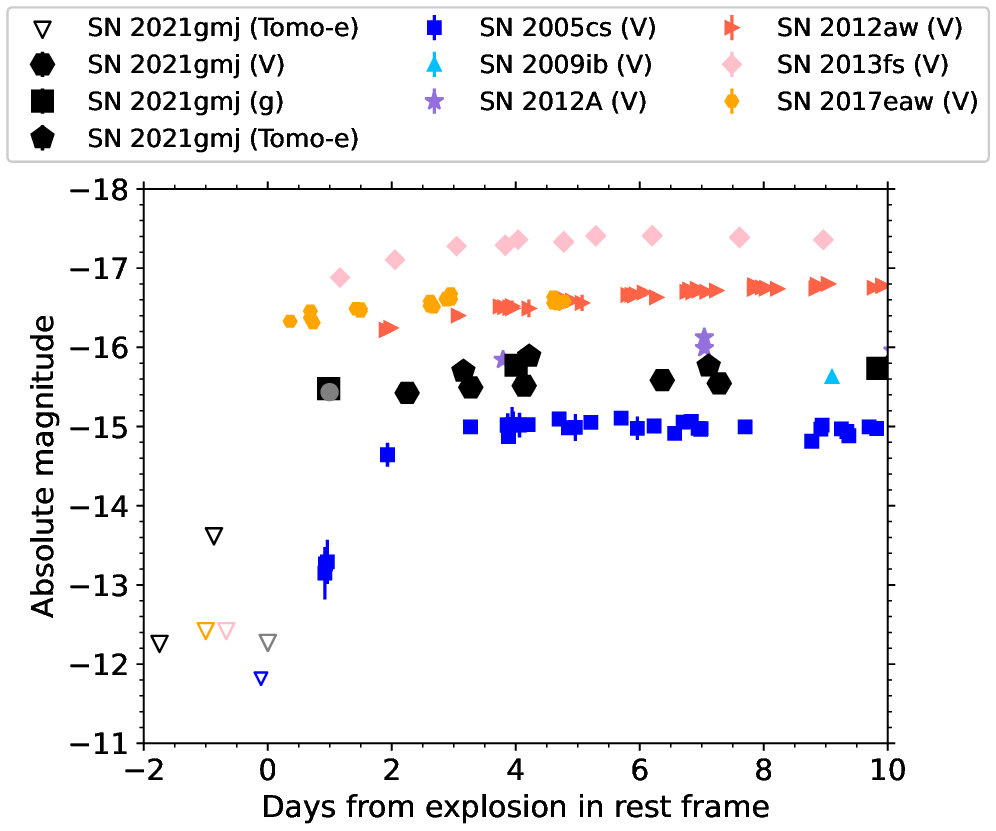}
\caption{
  \label{fig:early_compareMag}
  Early light curves of SN 2021gmj and other Type IIP SNe: 
  SN 2005cs (\citealt{Tsvetkov2006, Pastorello2006, Pastorello2009}), 
  SN 2009ib (\citealt{Takats2015}),  
  SN 2012A (\citealt{Tomasella2013}), SN 2012aw (\citealt{Fraser2012}), 
  SN 2013fs (\citealt{Valenti2016, Bullivant2018}) and 
  SN 2017eaw (\citealt{Tsvetkov2018}). 
  The Tomo-e, $V$-, and $g$-band data are plotted for SN 2021gmj, while the $V$-band data are plotted for the other SNe. 
  The inverse open triangles represent the upper limits. 
  The gray inverse open triangle and filled circle show the last non-detection and first detection for SN 2021gmj, respectively 
  (\citealt{Valenti2021}). 
  We obtained the data of the other SNe from the Open Supernova Catalog \citep{Guillochon2017}.
}
\end{center}
\end{figure}
%%%%%%%%%%%%%%%%%%%%%%%%%%%%%%%%%%%%%%%%%%%%%%%%%%%%

%%%%%%%%%%%%%%%%%%%%%%%%%%%%%%%%%%%%%%%%%%%%%%%%%%%% 
% Figure: Color
%%%%%%%%%%%%%%%%%%%%%%%%%%%%%%%%%%%%%%%%%%%%%%%%%%%% 
\begin{figure}
  \begin{center}
    \includegraphics[width=0.47\textwidth]{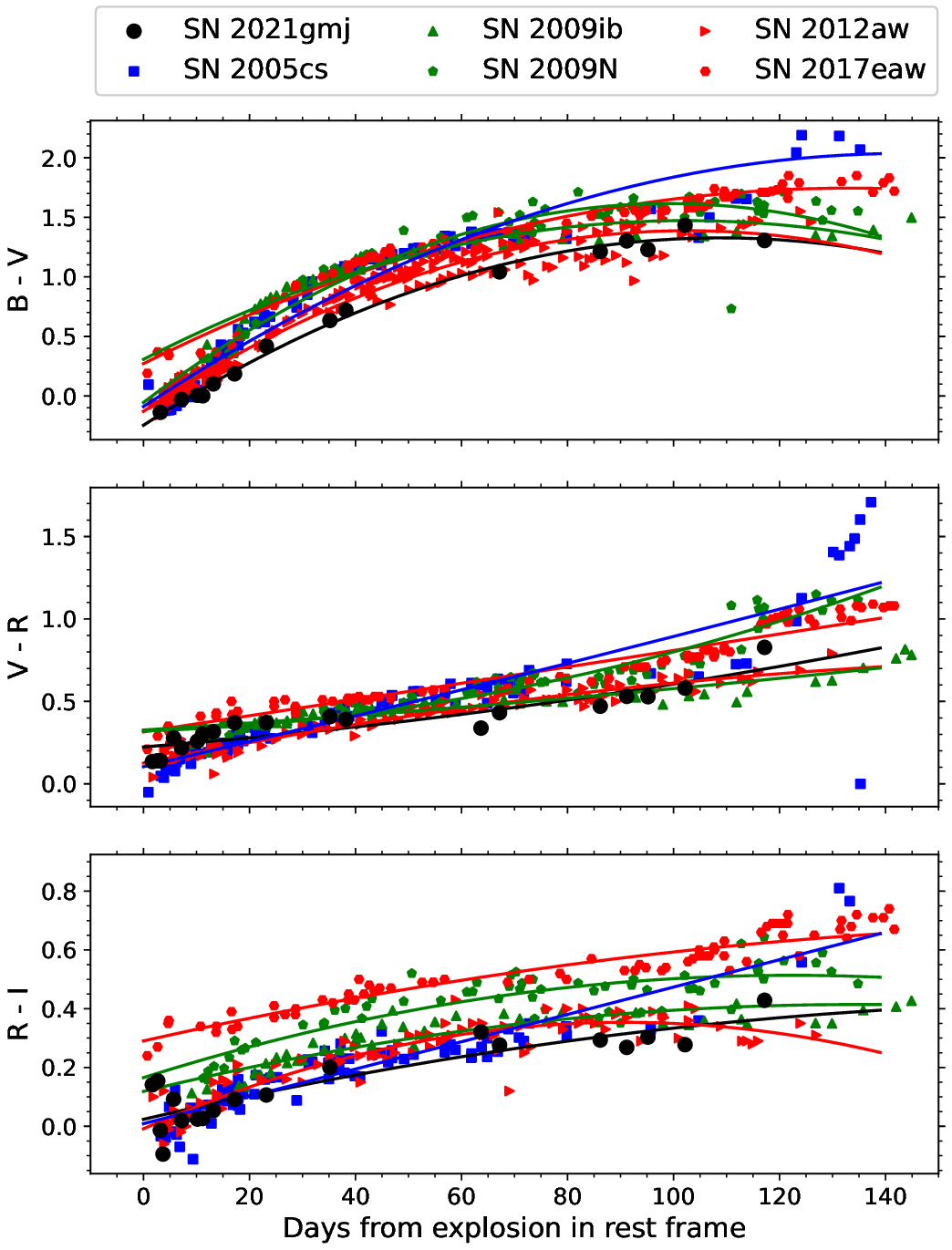}
  \caption{
  \label{fig:color}
  Color evolution of SN 2021gmj compared with other Type IIP SNe: the sub-luminous SN 2005cs (\citealt{Tsvetkov2006, Pastorello2006, Pastorello2009}), 
  the intermediate-luminosity SNe 2009ib (\citealt{Takats2015}), 
  2009N (\citealt{Takats2014}), 
  and the normal SNe 2012aw (\citealt{Fraser2012}), 
  2017eaw (\citealt{Tsvetkov2018}). 
  The data of the other SNe were obtained from the Open Supernova Catalog. 
  Red, green, and blue lines show the results of second order polynomial fitting for normal, intermediate-luminosity, and sub-luminous SNe, respectively.
}
\end{center}
\end{figure}
%%%%%%%%%%%%%%%%%%%%%%%%%%%%%%%%%%%%%%%%%%%%%%%%%%%%

Figure \ref{fig:magnitude} shows the multi-band light curves of SN 2021gmj. 
In addition to our data, we obtained $g$- and $r$-band data from 
Zwicky Transient Facility (ZTF; \citealt{Bellm2019}) taken through 
the Automatic Learning for the Rapid Classification of Events (ALeRCE; \citealt{Forster2021}). 
We estimate the explosion date of SN 2021gmj as $t_{0}$ = 59292.321 MJD 
from a comparison of the Tomo-e- and $r$-band light curves with light curve models 
of \cite{Moriya2023} (see Section \ref{subsec:properties}). 
For the other objects compared in following figures, 
we adopt the explosion time estimated in each object's literature reference:
they are typically estimated by the last non-detection and discovery dates 
and/or by analytic fitting of the rising part of the light curves.

SN 2021gmj reaches its brightness peak in optical bands very quickly (4-6 days) after the explosion. 
The absolute peak magnitudes of SN 2021gmj in the $V$ and $R$ bands are 
$\sim-15.5$ mag and $\sim-15.9$ mag, respectively, at $\sim7$ days after the explosion. 
During the plateau phase, 
the Tomo-e, $V$-, $R$-, $r$- and $I$-band magnitudes are roughly constant, 
the $J$-, $H$- and $K_{s}$-band magnitudes slightly increase for $\sim70$ days, 
and the $B$- and $g$-band magnitudes decline steadily. 
The plateau length of SN 2021gmj from the peak measured in the $V$ band is about 100 days, 
which is consistent with a typical plateau length of Type II SNe 
($\sim80$-120 days; \citealt{Valenti2016}). 
Type II SNe have a diversity in the plateau absolute magnitude
($-18 \lesssim M_{V,\rm peak} \lesssim -14$) 
with a typical value of $\sim -16.7$ mag
(\citealt{Anderson2014, Valenti2016}). 
The plateau absolute magnitude of SN 2021gmj (about $-15.5$ mag in the $V$ band) 
indicates the nature of intermediate-luminosity. 
In this paper, we define SNe with $-16 \lesssim M_{V,\rm peak} \lesssim -15$ as intermediate-luminosity SNe, and SNe fainter than intermediate-luminosity SNe as sub-luminous SNe. 
Even if we consider the maximum extinction of $A_{V} \sim 0.3$ mag, the absolute magnitude of SN 2021gmj
still falls in the intermediate luminosity range.

We compare the light curves of SN 2021gmj with those of other Type IIP SNe 
in Figure \ref{fig:compareMag}. 
The comparison SNe are the sub-luminous SN 2005cs 
(\citealt{Tsvetkov2006}; \citealt{Pastorello2006}; \citealt{Pastorello2009}), 
the normal SNe 2012aw (\citealt{Fraser2012}), 
2013fs (\citealt{Valenti2016}; \citealt{Bullivant2018}), and 2017eaw (\citealt{Tsvetkov2018}) 
and the intermediate-luminosity SNe 2009ib (\citealt{Takats2015}), 2009N (\citealt{Takats2014}), 
and 2012A (\citealt{Tomasella2013}). 
The plateau length of SN 2021gmj ($\sim100$ days) is similar to those of SNe 2009N and 2012A, 
and shorter than those of SNe 2005cs, 2009ib, 2012aw, and 2017eaw. 
While there is not sufficient data of SN 2013fs after the plateau phase, 
the plateau length of SN 2021gmj is longer than that of SN 2013fs. 

The brightness drop from the plateau to tail phases in SN 2021gmj is $\sim2$ mag 
with a duration of $\sim20$ days in the $V$ band. 
This brightness drop is smaller than that of SN 2005cs, 
while it is similar to those of the other objects. 
After this drop in luminosity, 
the light curve shows a linear decline powered by the radioactive decay of \Cofs~to \Fefs. 
In the tail phase, SN 2021gmj is brighter than SNe 2005cs and 2012A while fainter than SN 2009ib. 
The luminosity in the tail phase of SN 2021gmj is similar to that of SN 2009N. 
We estimate the \Nifs ~mass of SN 2021gmj from the luminosity in the tail phase 
in Section \ref{subsec:bolometricLC}.

The early-phase light curves of SN 2021gmj and other Type IIP SNe are shown in Figure \ref{fig:early_compareMag}. 
For the other Type IIP SNe shown in the plot, 
we adopt the explosion date estimated in each object's literature reference.
Although there is no data during the rising part of SN 2021gmj, 
our observations combined with the ZTF data show that
the duration of the rising part in SN 2021gmj resembles those of the other SNe.

The $B-V$, $V-R$ and $R-I$ color evolutions of SN 2021gmj are shown in Figure \ref{fig:color}. 
The colors are compared to those of the sub-luminous SN, the intermediate-luminosity SNe and the normal SNe that are used for the photometric comparison in Figure \ref{fig:compareMag}. 
The solid curves show the fitting by the second order polynomial. The overall color evolutions of SN 2021gmj are within the range of other Type IIP SNe.

%%%%%%%%%%%%%%%%%%%%%%%%%%%%%%%%%%%%%%%%%%%%%%%%%%%% 
% Figure: Spectra
%%%%%%%%%%%%%%%%%%%%%%%%%%%%%%%%%%%%%%%%%%%%%%%%%%%% 
\begin{figure*}
  \begin{center}
    \includegraphics[width=\textwidth]{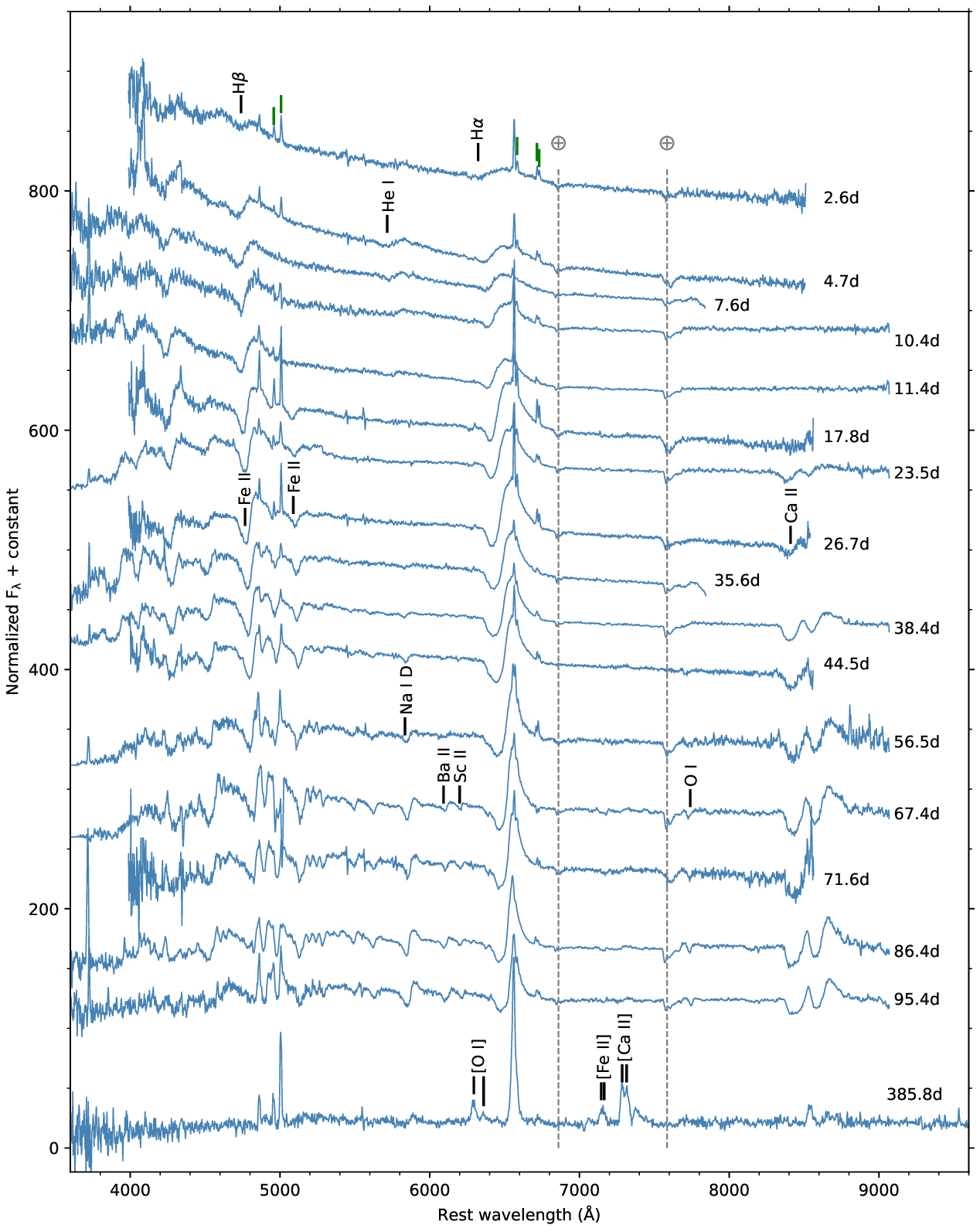}
\caption{
  \label{fig:spectra}
  Spectral evolution of SN 2021gmj from 2.6 days to 385.8 days after the explosion. 
  The green lines represent [O III] $\rm \lambda$4959,5007, [N II] $\rm \lambda$6584 and [S II] $\rm \lambda$6716,6731 
  originating from the nearby H II region.
  The narrow Balmer lines are also due to the contamination from the nearby H II region.
}
\end{center}
\end{figure*}
%%%%%%%%%%%%%%%%%%%%%%%%%%%%%%%%%%%%%%%%%%%%%%%%%%%%

%%%%%%%%%%%%%%%%%%%%%%%%%%%%%%%%%%%%%%%%%%%%%%%%%%%% 
% Figure: Compare spectra
%%%%%%%%%%%%%%%%%%%%%%%%%%%%%%%%%%%%%%%%%%%%%%%%%%%% 
\begin{figure*}
  \begin{center}
    \includegraphics[width=\textwidth]{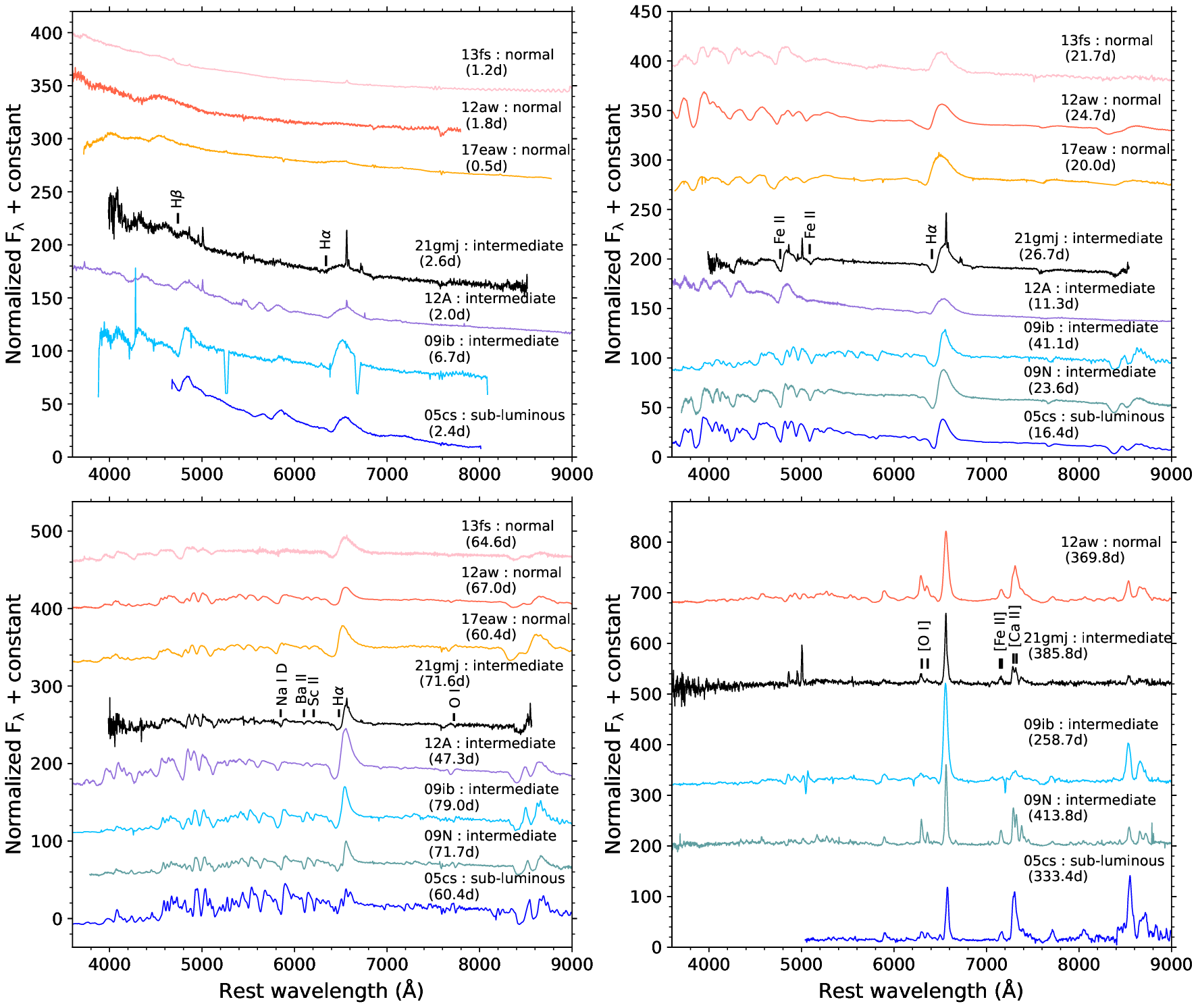}
\caption{
  \label{fig:compareSpectra}
  Spectra of SN 2021gmj from 2.6 to 385.8 days after the explosion, 
  compared with the sub-luminous SN 2005cs (\citealt{Tsvetkov2006, Pastorello2006, Pastorello2009}), 
  the intermediate-luminosity SNe 2009ib (\citealt{Takats2015}), 
  2009N (\citealt{Takats2014}) and 2012A (\citealt{Tomasella2013}), 
  and the normal SNe 2012aw (\citealt{Fraser2012}), 2013fs (\citealt{Valenti2016, Bullivant2018}) 
  and 2017eaw (\citealt{Szalai2019, VanDyk2019}). 
  We obtained the spectra of these Type IIP SNe from the Weizmann Interactive Supernova Data Repository (WISeREP; \url{https://www.wiserep.org/}; \protect\citealt{Yaron2012}). 
}
\end{center}
\end{figure*}
%%%%%%%%%%%%%%%%%%%%%%%%%%%%%%%%%%%%%%%%%%%%%%%%%%%%
%%%%%%%%%%%%%%%%%%%%%%%%%%%%%%%%%%%%%%%%%%%%%%%%%%%% 
% Figure: Velocity
%%%%%%%%%%%%%%%%%%%%%%%%%%%%%%%%%%%%%%%%%%%%%%%%%%%% 
\begin{figure}
  \begin{center}
    \includegraphics[width=0.47\textwidth]{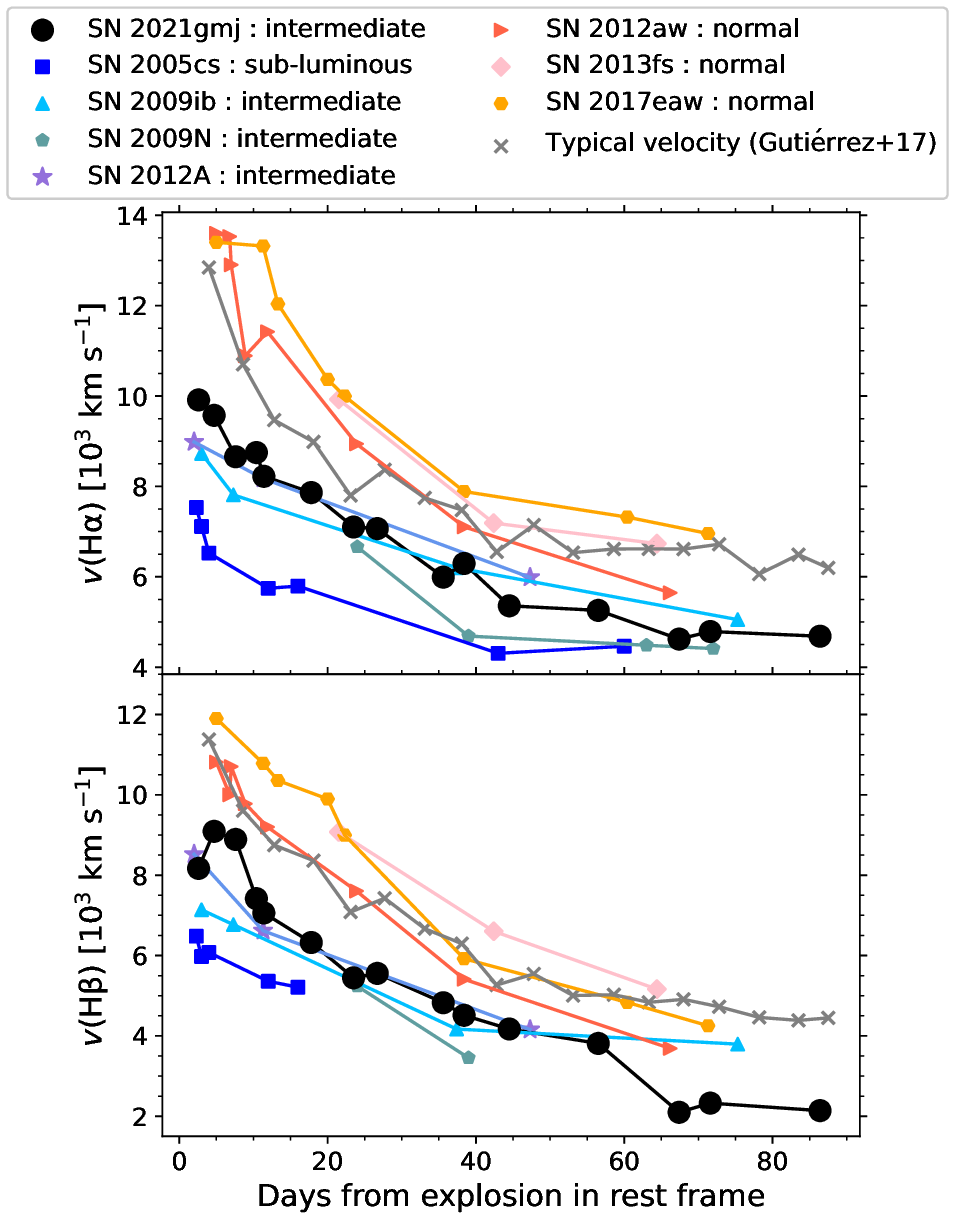}
\caption{
  \label{fig:velocity}
  Velocity evolution of H$\rm \alpha$ and H$\rm \beta$ lines in SN 2021gmj compared with those of other Type IIP SNe. 
  The gray lines are the results of \protect\cite{Gutierrez2017}, that are the typical velocities for Type II SNe.
  }
\end{center}
\end{figure}
%%%%%%%%%%%%%%%%%%%%%%%%%%%%%%%%%%%%%%%%%%%%%%%%%%%%
%%%%%%%%%%%%%%%%%%%%%%%%%%%%%%%%%%%%%%%%%%%%%%%%%%%% 
% Figure: Compare [Ca II]
%%%%%%%%%%%%%%%%%%%%%%%%%%%%%%%%%%%%%%%%%%%%%%%%%%%% 
\begin{figure}
  \begin{center}
    \includegraphics[width=0.47\textwidth]{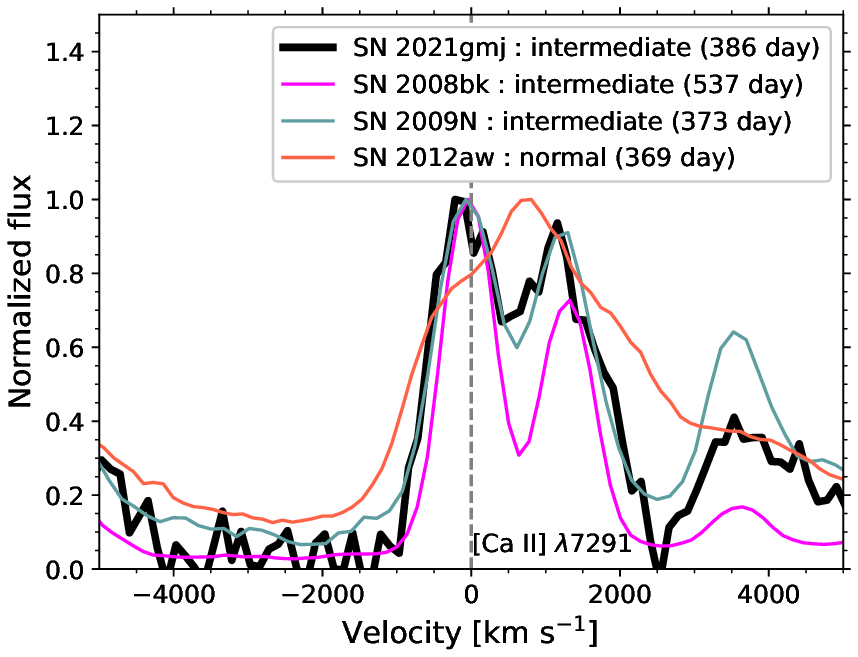}
\caption{
  \label{fig:compareCaII}
  Line profiles of the [Ca II] $\lambda$7291,7324 lines in the nebular spectrum of SN 2021gmj ($R \sim 360$) 
  compared with those of SNe 2008bk ($R \sim 730$), 2009N ($R \sim 520$), and 2012aw ($R \sim 460$). 
  The [Ca II] lines are doublet, and the velocity is measured with respect to the rest wavelength of the bluer feature.
  We obtained the data from WISeREP.
}
\end{center}
\end{figure}
%%%%%%%%%%%%%%%%%%%%%%%%%%%%%%%%%%%%%%%%%%%%%%%%%%%%

\subsection{Spectral evolution}
\label{subsec:spectroevolution}

The time series of the spectra of SN 2021gmj is presented in Figure \ref{fig:spectra}.
Overall, the spectra of SN 2021gmj show similar temporal evolution to other Type IIP SNe.
As in other Type IIP SNe, the early spectra (< 5 days) are characterized 
by broad Balmer and He I $\rm \lambda$5876 lines superposed on a blue continuum.
The H absorption lines generally get stronger as time goes by 
while the He I line disappears after 17.8 days after the explosion.
The metal lines such as Fe II $\rm \lambda$4924, 5169, and Ca II IR triplet 
start to become prominent on 26.7 days after the explosion. 
In the latter half of the plateau phase (from day 56.5 to 95.4), 
the metal lines become stronger 
and the lines of Na I D $\rm \lambda$5889, 5895 and O I $\rm \lambda$7774 are visible. 
The Ba II $\rm \lambda$6142 and Sc II $\rm \lambda$6247 lines, 
as also seen in other SNe at similar epochs 
(\citealt{Pastorello2004}; \citealt{Spiro2014}; \citealt{Gutierrez2017}; \citealt{Pearson2023}), 
are also visible. 
In the last spectrum in the nebular phase, several typical emission lines are detected such as 
[O I] $\rm \lambda$6300, 6364, [Ca II] $\rm \lambda$7291, 7324, and [Fe II] $\rm \lambda$7155, 7172. 

In Figure \ref{fig:compareSpectra}, we compare the spectra of SN 2021gmj with those of the sub-luminous SN 2005cs, 
the intermediate-luminosity SNe 2009ib, 2009N, and 2012A, 
and the normal SNe 2012aw, 2013fs, and 2017eaw. 
Overall, the spectra of SN 2021gmj have a similarity to those of the intermediate-luminosity SNe and sub-luminous SNe rather than those of the normal SNe.
In the earliest phase ($<$ 5 days, left upper panel of Figure \ref{fig:compareSpectra}), 
the spectrum of SN 2021gmj shows more structure than the normal SNe, 
and it is similar to those of the intermediate-luminosity and sub-luminous SNe. 

During the plateau phase (right upper and left lower panels of Figure \ref{fig:compareSpectra}), 
the absorption features in SN 2021gmj are narrower than those in normal SNe, 
and similar to those in intermediate-luminosity and sub-luminous SNe.
We estimate the velocities of H$\rm \alpha$ and H$\rm \beta$ lines 
by measuring their absorption minima after smoothing the spectra. 
The derived velocity evolution is shown in Figure \ref{fig:velocity} 
as compared with those of other Type IIP SNe. 
SN 2021gmj shows higher velocities than the sub-luminous SN 2005cs, 
while SN 2021gmj shows lower velocities compared to the normal SNe 2012aw, 2013fs, and 2017eaw. 
The velocities of SN 2021gmj are indeed similar to 
those of the intermediate-luminosity SNe 2009ib, 2009N, and 2012A. 

A statistical study about ejecta expansion velocities for Type II SNe (\citealt{Gutierrez2017}) shows that the range in ejecta expansion velocities 
is from $\sim1,500 ~ \rm km ~ s^{-1}$ to $\sim 9,600 ~ \rm km ~ s^{-1}$ 
at 50 days after explosion with a median H$\rm \alpha$ value of 7,300 
$\rm km ~ s^{-1}$. The H$\rm \alpha$ velocity of SN 2021gmj at 50 days after 
explosion is $\sim5,300 ~ \rm km ~ s^{-1}$, indicating SN 2021gmj is a low-velocity SN.

The low velocities of SN 2021gmj persist in the nebular phase 
(right lower panel of Figure \ref{fig:compareSpectra}).
We compare the line profile of [Ca II] $\lambda$7291,7324 
with those of the intermediate-luminosity SNe 2008bk, 2009N, and normal SN 2012aw 
in Figure \ref{fig:compareCaII}. 
The line width of SN 2021gmj is consistent with the spectral resolution ($R\sim 360$), indicating that the velocity of SN 2021gmj is lower than $\sim 1,000 ~ \rm km ~ s^{-1}$. 
SN 2021gmj clearly shows a narrower line profile than that of the normal SN 2012aw. 
The velocity of SN 2021gmj in the nebular phase is consistent with those of the intermediate-luminosity SNe 2008bk and 2009N. 

In summary, the spectral properties of SN 2021gmj are quite similar to 
those of the intermediate-luminosity SNe 2009ib, 2009N, and 2012A. 
This indicates that SN 2021gmj belongs to a population between normal SNe and sub-luminous SNe in terms of spectroscopic properties.

%%%%%%%%%%%%%%%%%%%%%%%%%%%%%%%%%%%%%%%%%%%%%%%%%%%% 
% Figure: Polarimetry
%%%%%%%%%%%%%%%%%%%%%%%%%%%%%%%%%%%%%%%%%%%%%%%%%%%% 
\begin{figure}
  \begin{center}
    \includegraphics[width=0.47\textwidth]{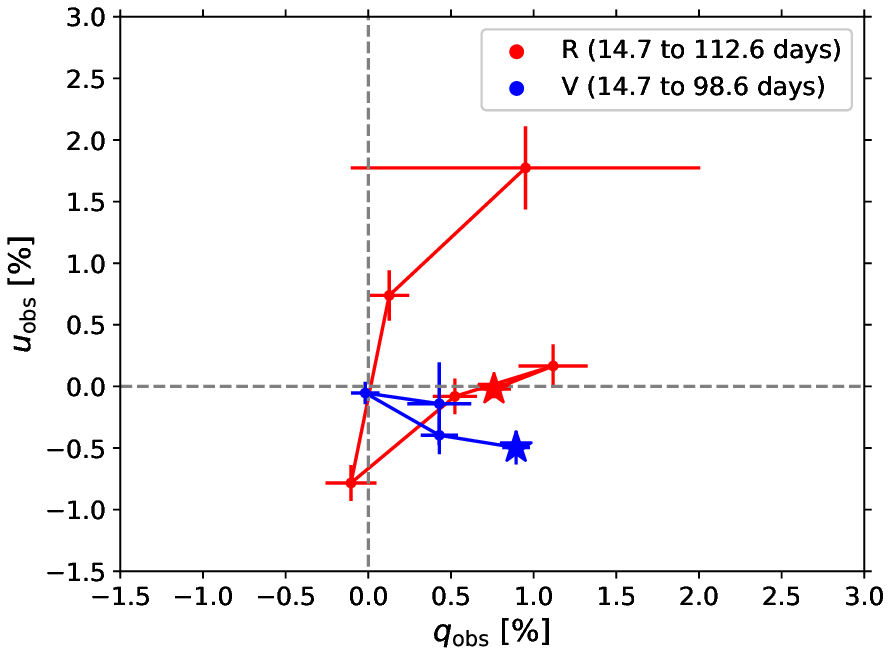}
\caption{
  \label{fig:resultP}
    The observed Stokes parameters in the $q-u$ plane. 
    The first epoch in both bands is 14.7 days after the explosion. 
    The final epoch in $V,R$ bands is 98.6, 112.6 days, respectively. 
    The stars represent the values in the first epoch.
}
\end{center}
\end{figure}
%%%%%%%%%%%%%%%%%%%%%%%%%%%%%%%%%%%%%%%%%%%%%%%%%%%%

\subsection{Polarimetric evolution}
\label{subsec:polarimetricevolution}
We present imaging polarimetry of SN 2021gmj obtained over six epochs 
from $\sim14$ to 112 days after the explosion. 
The observed Stokes parameters ($q_{\rm obs}$, $u_{\rm obs}$) 
in the $V$ and $R$ bands are shown in Figure \ref{fig:resultP}. 
During the plateau phase until $105.4\pm1.4$ days after explosion, 
the Stokes parameters are clustered 
around a point ($q\sim0.5$ \% and $u\sim-0.2$ \%) in the $q-u$ plane, and they 
do not significantly vary during the plateau phase. 
SN~2021gmj then started showing a deviation from this point in the $q-u$ plane 
at the beginning of the tail phase ($2.0\pm0.7$ days after the end of the plateau phase). 

In addition to the intrinsic SN polarization, 
the SN light would also be polarized by non-spherical dust grains 
in the interstellar medium along the line of sight. 
This is called interstellar polarization (ISP). 
In order to investigate the intrinsic property of the SN polarization, 
it is necessary to remove the ISP component. 
For an accurate estimate of ISP, one must know the properties of dust grain  along the line of sight 
within the host galaxy, which is usually difficult to obtain. 
Although many methods have been suggested to derive ISP (\citealt{Trammell1993, Tran1997, Wang2001}), 
each method involves some uncertainties (\citealt{WangWheeler2008}). 

Here we give only a conservative estimate of the ISP and intrinsic SN polarization. 
The degree of the ISP is observationally related to the degree of the interstellar extinction 
\citep[$p_{\rm ISP}\lesssim9\times E(B-V)$ \%,][]{Serkowski1975}. 
Non-detection of significant Na I D absorption lines in our spectra suggests an upper limit of the equivalent width of Na I D absorption lines is 0.7496 {\AA}.
This corresponds to $E(B-V) < 0.1$ by using the empirical relation by \cite{Poznanski2012}.
This gives an upper limit of the ISP $p_{\rm ISP}\lesssim0.9$ \%.
Thus, the obtained polarization degree in the plateau phase can be entirely caused by the ISP.
This implies that H-rich envelope of SN 2021gmj is not strongly deviated from spherical symmetry.
In general, Type IIP SNe show a low degree of the intrinsic polarization at early phases and 
show a rise to $\sim 1$ \% degree at certain points during the plateau phase 
(e.g., \citealt{Leonard2001, Chornock2010, Kumar2016, Nagao2023_1, Nagao2023_2}). 
The observed properties of SN 2021gmj are consistent with this picture, although it is difficult to quantitatively estimate the degree of the intrinsic polarization due to the uncertainty in the ISP.

%%%%%%%%%%%%%%%%%%%%%%%%%%%%%%%%%%%%%%%%%%%%%%%%%%%% 
% Section: Discussion
%%%%%%%%%%%%%%%%%%%%%%%%%%%%%%%%%%%%%%%%%%%%%%%%%%%% 

\section{Discussion}
\label{sec:discussion}

\subsection{Bolometric light curve}
\label{subsec:bolometricLC}

\begin{figure}
  \begin{center}
    \includegraphics[width=0.47\textwidth]{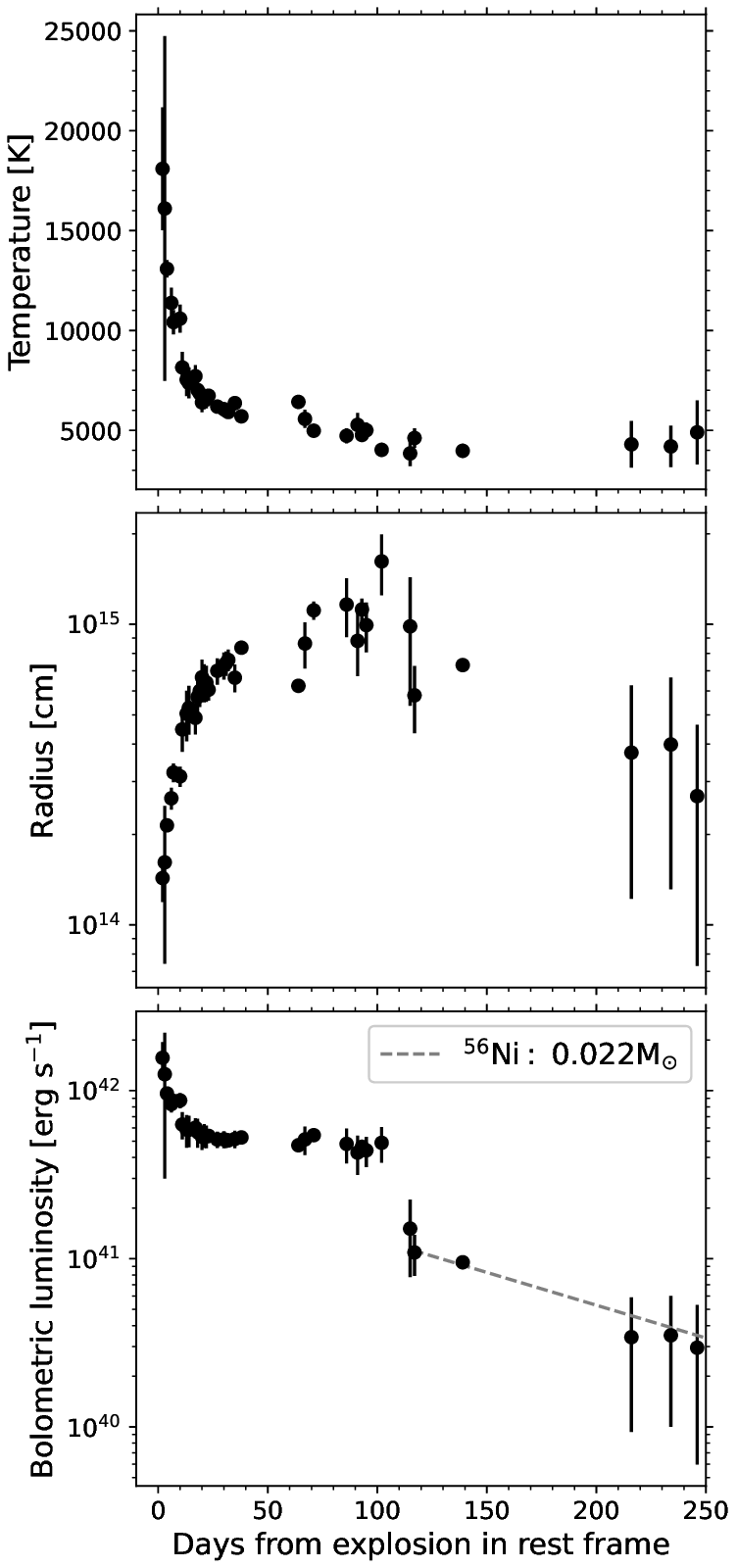}
\caption{
  \label{fig:bolometric}
  Blackbody evolutions of temperature (top), radius (middle), and bolometric luminosity (bottom) of SN 2021gmj. 
  The gray dashed line in the bottom figure represents the decay luminosity from the $\rm ^{56}Co$ decay (0.022 $\rm \Msun$).
}
\end{center}
\end{figure}

We derive the bolometric luminosity of SN 2021gmj 
by integrating the $U$-, $B$-, $V$-, $R$- and $I$-band photometry.  
For each observational epoch with more than 3 band data, 
we derived the fluxes at the effective wavelengths to specify the spectral energy distribution (SED) 
for each epoch. 
Then, we performed a blackbody fitting to the SED and derived the best-fit temperature and radius.
Finally, the bolometric luminosity was derived 
by integrating the best-fit blackbody function as shown in Figure \ref{fig:bolometric}. 

In Type IIP SNe, the luminosity at the tail phase is mainly powered 
by the radioactive decay of \Cofs. 
Therefore, the bolometric luminosity at the tail phase is a good indicator of the amount of \Nifs~
in the SN ejecta.
This luminosity can be described as follows: \\
\begin{equation}
\label{eq:decay}
L_{\rm decay}(t) = (6.45 \times 10^{43} e^{-t/\tau_{\rm Ni}} ~ + ~ 1.45 \times 10^{43} e^{-t/\tau_{\rm Co}}) \left(\frac{M_{\rm Ni}}{\Msun} \right) ~ \rm erg ~ s^{-1} ~ ,
\end{equation}
where $t$ is time in days (\citealt{Nadyozhin1994}). 
$\tau_{\rm Ni}$ and $\tau_{\rm Co}$ are the lifetime of \Nifs~ and \Cofs~ in days, respectively. 
From this equation and the tail phase of the bolometric light curve of SN 2021gmj, 
the \Nifs~mass for SN 2021gmj is estimated to be $0.022\pm0.006 ~ \rm \Msun$ (Figure \ref{fig:bolometric}). 

The \Nifs~mass derived for SN 2021gmj is smaller than those of 
the normal SN 2012aw: $M(^{56}{\rm Ni}) = 0.05 ~ \rm \Msun$ (\citealt{NagyVinko2016}), 
the normal SN 2017eaw: $M(^{56}{\rm Ni}) = 0.046 ~ \rm \Msun$ (\citealt{Szalai2019}), and 
the intermediate-luminosity SN 2009ib: $M(^{56}{\rm Ni}) = 0.046\pm0.015 ~ \rm \Msun$ (\citealt{Takats2015}). 
On the other hand, the \Nifs ~mass of SN 2021gmj is larger than those of the sub-luminous SN 2005cs: 
$M(^{56}{\rm Ni}) \sim 0.003 ~ \rm \Msun$ (\citealt{Pastorello2009}) 
and the intermediate-luminosity SN 2012A: 
$M(^{56}{\rm Ni}) = 0.011\pm0.004 ~ \rm \Msun$ (\citealt{Tomasella2013}). 
The \Nifs ~mass in SN 2021gmj is the most similar to that in the intermediate-luminosity SN 2009N: 
$M(^{56}{\rm Ni}) = 0.020 \pm 0.004 ~ \rm \Msun$ (\citealt{Takats2014}). 

The \Nifs~mass in SN 2021gmj can also be compared with the results of statistical studies of Type II SNe. For instance, \cite{Muller2017} investigated the observed distribution of \Nifs~mass for 38 Type II SNe, 
showing a median and mean values of 0.031 and 0.046 $\rm \Msun$, respectively. \cite{Anderson2019} estimated a median \Nifs~mass for 115 Type II SNe to be 0.032 $\rm \Msun$. \cite{Rodriguez2021} investigated the \Nifs~mass distribution for 110 Type II SNe with the average of $0.040 \pm 0.005 ~ \rm \Msun$. More recently, \cite{Martinez2022} determined a median \Nifs~mass of 0.036 $\rm \Msun$ for 17 Type II SNe, that have the \Nifs~mass distribution between 0.006 and 0.069 $\rm \Msun$. 

%%%%%%%%%%%%%%%%%%%%%%%%%%%%%%%%%%%%%%%%%%%%%%%%%%%% 
% Figure: 56Ni (bolometric)
%%%%%%%%%%%%%%%%%%%%%%%%%%%%%%%%%%%%%%%%%%%%%%%%%%%% 
%%%%%%%%%%%%%%%%%%%%%%%%%%%%%%%%%%%%%%%%%%%%%%%%%%%%

\subsection{Progenitor mass}
\label{subsec:M_prog}

Spectroscopic observations of SNe at nebular phase provide useful information 
on the nucleosynthesis in SN explosions and their progenitors. 
As SN ejecta expands, their density decreases, and the inner parts are exposed. 
When the inner core becomes optically thin, 
the observed spectrum should be dominated by emission lines. 
In particular, the ratio of [O I] $\lambda$6300, 6364 and [Ca II] $\lambda$7291, 7324 lines is used as an indicator of progenitor mass. 
The [O I]/[Ca II] ratio is correlated with the progenitor oxygen mass, 
which is dependent on the progenitor mass 
(\citealt{FranssonChevalier1989, maeda2007, Jerkstrand2015, Kuncarayakti2015, Jerkstrand2017, Fang2022, Fang2023}). 

\begin{figure}
  \begin{center}
    \includegraphics[width=0.47\textwidth]{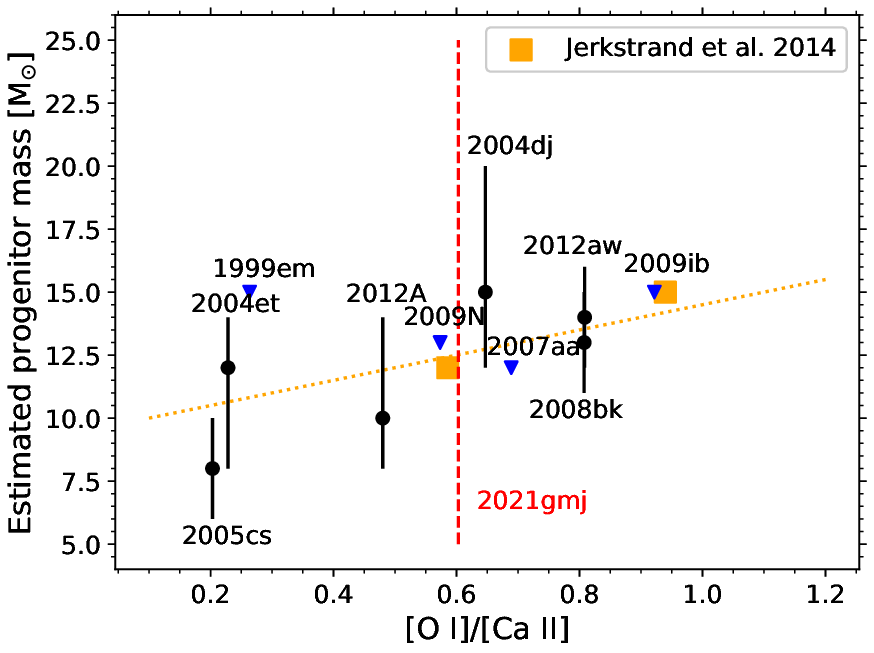}
\caption{
  \label{fig:OICaII}
  [O I]/[Ca II] ratio of SN 2021gmj and Type IIP SNe whose progenitor masses were estimated from the pre-explosion image analysis: 
  SN 1999em (\citealt{Kuncarayakti2015, Smartt2003, Smartt2009}), 
  SN 2004dj (\citealt{Chugai2005, Maiz-Apellaniz2004, Wang2005, Smartt2009}), 
  SN 2004et (\citealt{Kuncarayakti2015, Smartt2009}), 
  SN 2005cs (\citealt{Maund2005, Li2006, Pastorello2009, Smartt2009}), 
  SN 2007aa (\citealt{Smartt2009}), SN 2008bk (\citealt{Maund2014}), 
  SN 2009N (\citealt{Smartt2015}), SN 2009ib (\citealt{Takats2015, Smartt2015}), 
  SN 2012A (\citealt{Tomasella2013, Smartt2015}), and 
  SN 2012aw (\citealt{Fraser2012, Smartt2015}). 
  The object whose progenitor is detected in its pre-explosion image is represented in black dot 
  while the object whose progenitor is not detected is represented in blue inverse triangle as an upper limit. 
  The red dashed line represents the [O I]/[Ca II] ratio of SN 2021gmj. 
  The orange square dots show the relation between the progenitor mass 
  and [O I]/[Ca II] ratio of spectral models for Type IIP SNe (\citealt{Jerkstrand2014}).
}
\end{center}
\end{figure}

In the nebular spectrum of SN 2021gmj at 385.8 days after the explosion, 
we measured the line ratio of [O I]/[Ca II] to be 0.60 
by fitting the [O I] and [Ca II] lines with double Gaussian functions. 
In order to infer the progenitor mass, 
we compared the line ratio with those of other Type IIP SNe with good constraints to the progenitor's zero-age main-sequence (ZAMS) mass from their pre-explosion images: 
SN 1999em (\citealt{Kuncarayakti2015, Smartt2003, Smartt2009}), 
SN 2004dj (\citealt{Chugai2005, Maiz-Apellaniz2004, Wang2005, Smartt2009}), 
SN 2004et (\citealt{Kuncarayakti2015, Smartt2009}), 
SN 2005cs (\citealt{Maund2005, Li2006, Pastorello2009, Smartt2009}), 
SN 2007aa (\citealt{Smartt2009}), SN 2008bk (\citealt{Maund2014}), 
SN 2009N (\citealt{Smartt2015}), SN 2009ib (\citealt{Takats2015, Smartt2015}), 
SN 2012A (\citealt{Tomasella2013, Smartt2015}), 
and SN 2012aw (\citealt{Fraser2012, Smartt2015}). 
The line ratios of these SNe are also measured by fitting double Gaussian to the lines. 
As shown in Figure \ref{fig:OICaII}, the [O I]/[Ca II] ratio shows a weak linear correlation with the progenitor mass.
From this correlation, the progenitor mass of SN 2021gmj is estimated to be about 10 to 15 $\rm \Msun$.

\begin{figure*}
  \begin{center}
    \includegraphics[width=\textwidth]{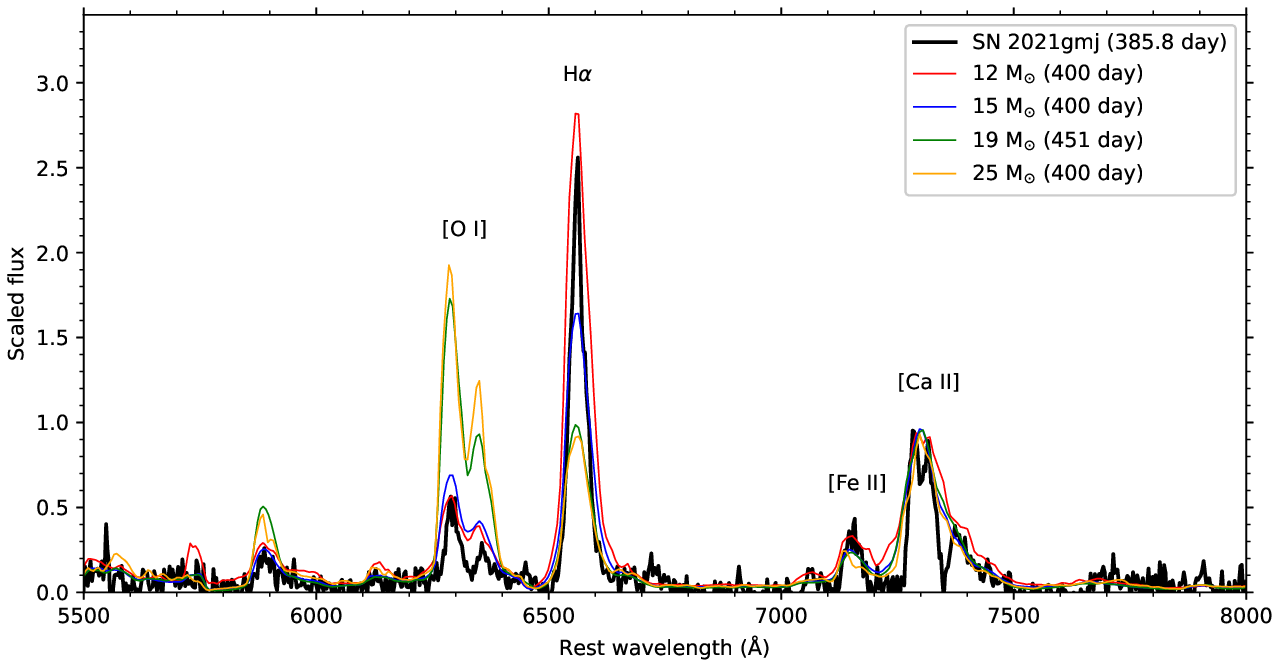}
\caption{
  \label{fig:spectralModeling}
  Comparison of the nebular spectrum of SN 2021gmj with the spectral models of \protect\cite{Jerkstrand2014}. 
  Spectra are scaled to match the [Ca II] $\lambda$7291,7324 flux. 
  The flux of narrow H$\rm \alpha$ from the host galaxy is subtracted from the observed flux 
  assuming the flux ratio H$\rm \alpha$/H$\rm \beta$ to be 3:1.
}
\end{center}
\end{figure*}

We also measure the [O I]/[Ca II] ratios in the spectral models for Type IIP SNe 
by \cite{Jerkstrand2014} (see Figure \ref{fig:OICaII}). 
\cite{Jerkstrand2014} generated models of nebular spectra 
from stellar evolution and explosion models
using the spectral synthesis code described in \cite{Jerkstrand2011}.
These model spectra are calculated for the stars with 12, 15, 19 and 25 $\rm \Msun$ initial mass.
In Figure \ref{fig:spectralModeling}, we compare the [O I]/[Ca II] ratios of 12 and 15 $\rm \Msun$ models at 400 days 
and 19 $\rm \Msun$ model at 451 days with SN 2021gmj at 385.8 days, where the fluxes are scaled to match the [Ca II] $\lambda$7291, 7324 lines. 
From the comparison between the [O I]/[Ca II] ratios of the spectral models and SN 2021gmj, 
the progenitor mass of SN 2021gmj is likely to be about 12 $\rm \Msun$, which is consistent with the mass estimated from the weak correlation 
between the [O I]/[Ca II] ratios and the progenitor mass of the other Type IIP SNe. The 12 $\rm \Msun$ model also gives the best match with the SN spectrum from the similarities of [O I], [Fe II] and H$\alpha$ lines.
%%%%%%%%%%%%%%%%%%%%%%%%%%%%%%%%%%%%%%%%%%%%%%%%%%%% 
% Figure: [O I]/[Ca II]
%%%%%%%%%%%%%%%%%%%%%%%%%%%%%%%%%%%%%%%%%%%%%%%%%%%% 
%%%%%%%%%%%%%%%%%%%%%%%%%%%%%%%%%%%%%%%%%%%%%%%%%%%%
%%%%%%%%%%%%%%%%%%%%%%%%%%%%%%%%%%%%%%%%%%%%%%%%%%%% 
% Figure: Spectral modeling
%%%%%%%%%%%%%%%%%%%%%%%%%%%%%%%%%%%%%%%%%%%%%%%%%%%% 
%%%%%%%%%%%%%%%%%%%%%%%%%%%%%%%%%%%%%%%%%%%%%%%%%%%%

\subsection{Explosion and circumstellar material properties}
\label{subsec:properties}

The rising part of light curves of Type IIP SNe can be used 
as a probe of circumstellar material (CSM) 
(e.g., \citealt{Moriya2011, Moriya2017, Moriya2018, Morozova2017, Morozova2018, Forster2018}). 
In order to estimate explosion and CSM properties, 
we compared the Tomo-e and $r$-band light curves with the light curve models by \cite{Moriya2023}. 
In these models, red supergiant (RSG) models in \cite{Sukhbold2016} are adopted as the progenitors for Type IIP SNe. 
The confined CSM is attached on the surface of the progenitors. 
For the calculations of the light curve models, the radiation hydrodynamics code $\rm STELLA$ 
(\citealt{Blinnikov1998, Blinnikov2000, Blinnikov2006}) is used. 
To initiate the explosion, thermal energy is injected just above a mass cut 
to trigger the explosion.

The mass cut of the progenitor is set at 1.4 $\rm \Msun$. 
The density structure of the CSM is assumed as in \cite{Moriya2017, Moriya2018},  \\
\begin{equation}
\rho_{\rm CSM}\left (r \right) = \frac{\dot{M}}{4\pi v_{\rm wind}\left (r \right)r^{2}} ~ ,
\end{equation}
where $r$ is the radius from the center of a progenitor, 
$\dot{M}$ is the mass-loss rate of a progenitor, 
and $v_{\rm wind}$ is the wind velocity. 
The wind velocity is expressed as a function of radius: 
\begin{equation}
v_{\rm wind}(r) = v_{0} + (v_{\infty} - v_{0}) \left(1 - \frac{R_{0}}{r} \right) ^{\beta} ~ ,
\end{equation}
where $v_{0}$ is an initial wind velocity at the surface of a progenitor, 
$v_{\infty}$ is a terminal wind velocity, $R_{0}$ is a progenitor radius, 
and $\beta$ is a wind structure parameter. 
Here, $v_{0}$ is the value that smoothly connects the CSM to the progenitor surface, 
and $v_{\infty}$ is assumed to be $\rm 10 ~ km ~ s^{-1}$ \citep{Moriya2023}. 
The parameters in the model sets are the ZAMS mass of the progenitor ($M_{\rm ZAMS}$), 
the \Nifs~mass ($M_{\rm Ni}$), the wind parameter ($\beta$), the CSM radius ($R_{\rm CSM}$), 
the explosion energy ($E$), and the mass-loss rate ($\dot{M}$). 

Since the explosion energy and the mass-loss rate 
mainly affect the rising part of light curves, we fix the other parameters as follows: $M_{\rm ZAMS} = 12 ~ \rm \Msun$, $M_{\rm Ni} = 0.02 ~ \rm \Msun$, $\beta = 2$, and $R_{\rm CSM} = 4 \times 10^{14} ~ \rm cm$. 
As RSGs are known to accelerate the wind slowly (e.g., \cite{Schroeder1985}), we fix $\beta$ to be 2. 
Note that it does not affect the early light curves much. 
We assume the breakout time when a photosphere reaches to the outer edge of the CSM is 20 days after explosion based on the observed light curve (Figure \ref{fig:magnitude}). 
From this assumption, we fix $R_{\rm CSM} = 4 \times 10^{14}$ cm here.
It also means that the CSM affects the light curves only until 20 days after explosion, thus the explosion energy can be determined by the light curves after this epoch. 
We then vary the explosion energy (0.2, 0.3, 0.5, 1.0, and 1.5$\times ~ 10^{51} \rm erg$) and the mass-loss rate 
($10^{-5.0}, ~ 10^{-4.5}, ~ 10^{-4.0}, ~ 10^{-3.5}, ~ 10^{-3.0}, ~ 10^{-2.5},$ and $10^{-2.0} ~ \rm M_{\odot} ~ yr^{-1}$). 
The light curve models with 0.5, 1.0, and 1.5$\times ~ 10^{51} \rm erg$ of the explosion energy are calculated in \cite{Moriya2023}. 
We additionally calculated the light curves for these low-energy models (0.2 and 0.3 $\times ~ 10^{51} \rm erg$) using $\rm STELLA$ code in the same way as \cite{Moriya2023}. 
We compare the Tomo-e and $r$-band light curves 
of the early phase ($\leq$ 40 days after the explosion) with the available light curve models from \cite{Moriya2023} and the newly calculated low-energy models. 
In the comparison, the explosion date is also parameterized with a range 
between the last non-detection date and the discovery date. 

\begin{figure}
  \begin{center}
    \includegraphics[width=0.47\textwidth]{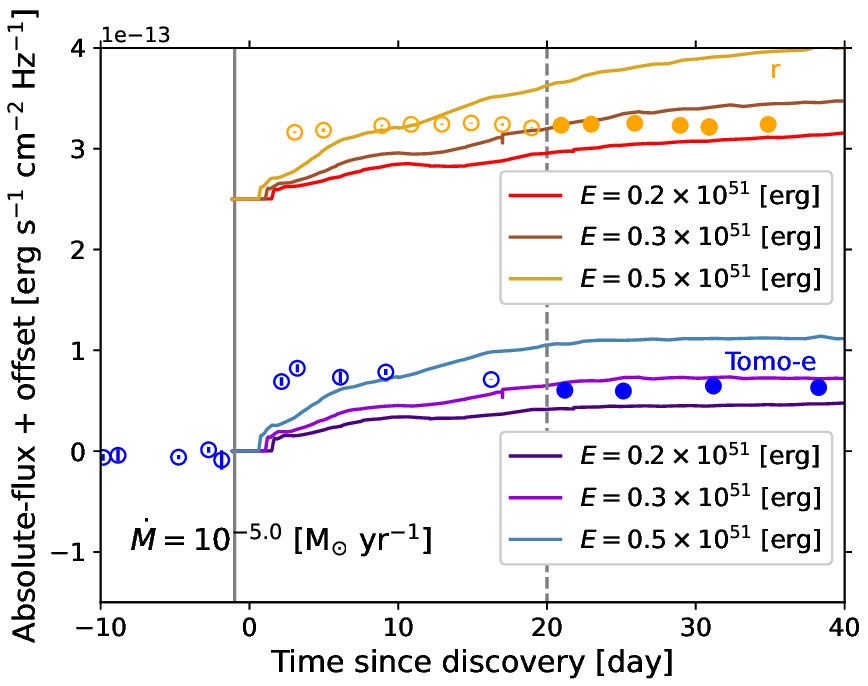}
\caption{
  \label{fig:chi2fit_E}
 Comparison of the observed light curves with the model light curves with $E=0.2, 0.3,$ and $0.5 \times 10^{51}$ erg. 
 Other parameters of the models are fixed: $M_{\rm ZAMS} = 12 ~ \rm \Msun$, $M_{\rm Ni} = 0.02 ~ \rm \Msun$, $\beta = 2$, $R_{\rm CSM} = 4 \times 10^{14} ~ \rm cm$ and $\dot{M} = 10^{-5} ~ \rm M_{\odot}~ yr^{-1}$. 
 The gray solid line is the estimated explosion date.
 The gray dashed line is the breakout time (20 days). 
 The model light curve with $E=0.5 \times 10^{51}$ erg of the explosion energy comes from \protect\cite{Moriya2023}. The model light curves with $E=0.2$ and $0.3\times 10^{51}$ erg of the explosion energy are calculated using STELLA code in the same way as \protect\cite{Moriya2023}.
}
\end{center}
\end{figure}

First, we estimate the explosion energy by comparing the light curve models with various explosion energy for a fixed mass-loss rate $10^{-5.0} ~ \rm M_{\odot} ~ yr^{-1}$ (Figure \ref{fig:chi2fit_E}).
For the estimate of the explosion energy, we use the plateau luminosity from the breakout time ($\sim 20$ days) to 40 days after the explosion. 
This is because: (a) plateau luminosity is mainly determined by explosion energy when a progenitor mass is fixed (e.g., \citealt{Kasen2009}); 
(b) the effect of CSM in light curves is not significant after the breakout time. 
Thus, this comparison is used only to estimate the explosion energy to reproduce the plateau luminosity. 
We find the model with $E = 0.3\times 10^{51}$ erg best matches to the light curves in both bands.

\begin{figure}
  \begin{center}
    \includegraphics[width=0.47\textwidth]{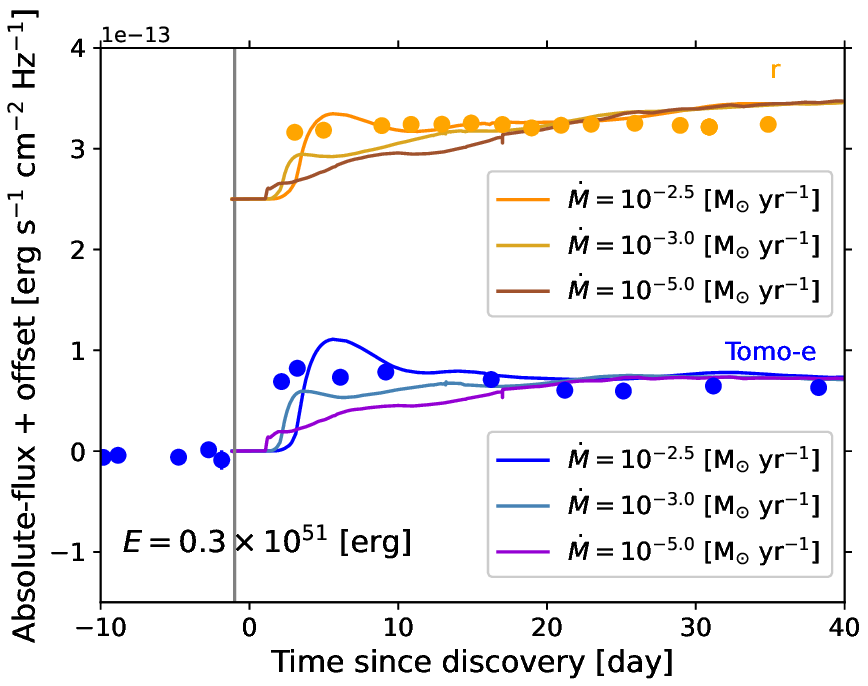}
\caption{
  \label{fig:chi2fit_Mdot}
 Comparison of the observed light curves with the model light curves with $\dot{M}=10^{-2.5},10^{-3},$ and $10^{-5}~\rm \Msun~yr^{-1}$. 
 Other parameters of the models are fixed: $M_{\rm ZAMS} = 12 ~ \rm \Msun$, $M_{\rm Ni} = 0.02 ~ \rm \Msun$, $\beta = 2$, $R_{\rm CSM} = 4 \times 10^{14} ~ \rm cm$ and $E = 0.3\times10^{51} ~ \rm erg$. The gray vertical line is the estimated explosion date.
}
\end{center}
\end{figure}

Then, with this estimated explosion energy, we estimate the mass-loss rate by changing the mass-loss rate (Figure \ref{fig:chi2fit_Mdot}). 
We find a higher mass-loss rate can reproduce the rising parts of the light curves 
better. 
As shown in Figure \ref{fig:chi2fit_Mdot}, 
the model with a high mass-loss rate ($\dot{M} = 10^{-3} ~ \rm \Msun ~ yr^{-1}$) 
reproduces the fast rise of the light curves 
as compared with the model with a normal mass-loss rate ($\dot{M} \lesssim 10^{-5} ~ \rm \Msun ~ yr^{-1}$; \citealt{Smith2014}), 
indicating that dense CSM exists around the progenitor of SN 2021gmj 
just before the explosion. 
This finding is consistent with those suggested for normal SNe (e.g., \citealt{Valenti2016, Morozova2017, Morozova2018, Forster2018}). 
%%%%%%%%%%%%%%%%%%%%%%%%%%%%%%%%%%%%%%%%%%%%%%%%%%%% 
% Figure: Model fitting
%%%%%%%%%%%%%%%%%%%%%%%%%%%%%%%%%%%%%%%%%%%%%%%%%%%% 
%%%%%%%%%%%%%%%%%%%%%%%%%%%%%%%%%%%%%%%%%%%%%%%%%%%%
%%%%%%%%%%%%%%%%%%%%%%%%%%%%%%%%%%%%%%%%%%%%%%%%%%%% 
% Figure: Model fitting
%%%%%%%%%%%%%%%%%%%%%%%%%%%%%%%%%%%%%%%%%%%%%%%%%%%% 
%%%%%%%%%%%%%%%%%%%%%%%%%%%%%%%%%%%%%%%%%%%%%%%%%%%%

\subsection{Early spectral features}
\label{subsec:earlyspec}

\begin{figure}
  \begin{center}
    \includegraphics[width=0.47\textwidth]{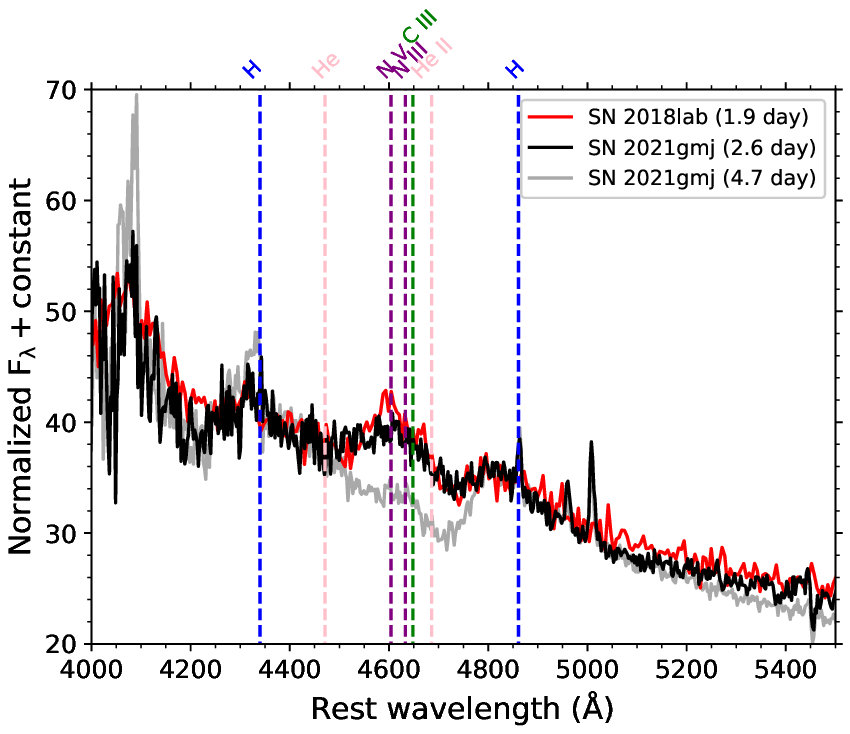}
\caption{
  \label{fig:compare_18lab}
 Comparison of the spectra of SN 2021gmj at 2.6 and 4.7 day with the spectrum of SN 2018lab at 1.9 day (\citealt{Pearson2023}). 
 The rest wavelengths of the candidate lines are shown by vertical dashed lines.
}
\end{center}
\end{figure}

Emission features in early spectra also provide clues to the properties of CSM. 
The spectrum of SN 2021gmj at 2.6 days exhibits a broad feature 
spreading from $\sim4500$ \AA~ to $\sim4750$ \AA ~(Figure \ref{fig:compare_18lab}). 
This broad feature around 4600 \AA ~is referred to as a "ledge" feature 
(\citealt{Andrews2019, Soumagnac2020, Hosseinzadeh2022, Pearson2023}), 
which is often seen in the early spectra of Type IIP SNe. 
This feature is also seen in the early spectra of some sub-luminous SNe: 
SN 2005cs (\citealt{Pastorello2006}), SN 2010id (\citealt{Gal-Yam2011}), 
SN 2016bkv (\citealt{Hosseinzadeh2018}), and SN 2018lab (\citealt{Pearson2023}). 

The origin of this ledge feature in early spectra of Type IIP SNe is explained 
with a blend of some ionized lines originating from their CSM 
(e.g., \citealt{Hosseinzadeh2022, 2020jfoTeja, Pearson2023}). 
We compare the first-epoch spectrum of SN 2021gmj with a spectrum of SN 2018lab 
at a similar epoch in Figure \ref{fig:compare_18lab}. 
There is a remarkable similarity in the overall spectral shapes of these objects. 
In the case of SN 2018lab, 
although this emission feature can be interpreted as high-velocity H$\rm \beta$, 
there is no indication of a high-velocity H$\rm \alpha$ emission at the same epoch.
Thus, it is concluded that this feature is due to the several ionized lines from the CSM: 
He II, C III, N III and N V (\citealt{Pearson2023}). 
In the case of SN 2021gmj, 
there is also no indication of a high-velocity feature of H$\rm \alpha$ at the same epoch.
Hence, we attribute these emission lines in SN 2021gmj to 
He II, C III, N III and N V lines originating from its CSM as in SN 2018lab. 
The fact that this feature disappeared in the spectrum at 4.7 days after the explosion 
supports this interpretation. 
This also implies the presence of a large amount of CSM around the progenitor of SN 2021gmj
as suggested from the early light curves.

%%%%%%%%%%%%%%%%%%%%%%%%%%%%%%%%%%%%%%%%%%%%%%%%%%%% 
% Figure: Compare with 2018lab
%%%%%%%%%%%%%%%%%%%%%%%%%%%%%%%%%%%%%%%%%%%%%%%%%%%% 
%%%%%%%%%%%%%%%%%%%%%%%%%%%%%%%%%%%%%%%%%%%%%%%%%%%%

\subsection{Comparison with other Type IIP SNe}

\label{subsec:IIPproperties}
\begin{figure}
  \begin{center}
    \includegraphics[width=0.47\textwidth]{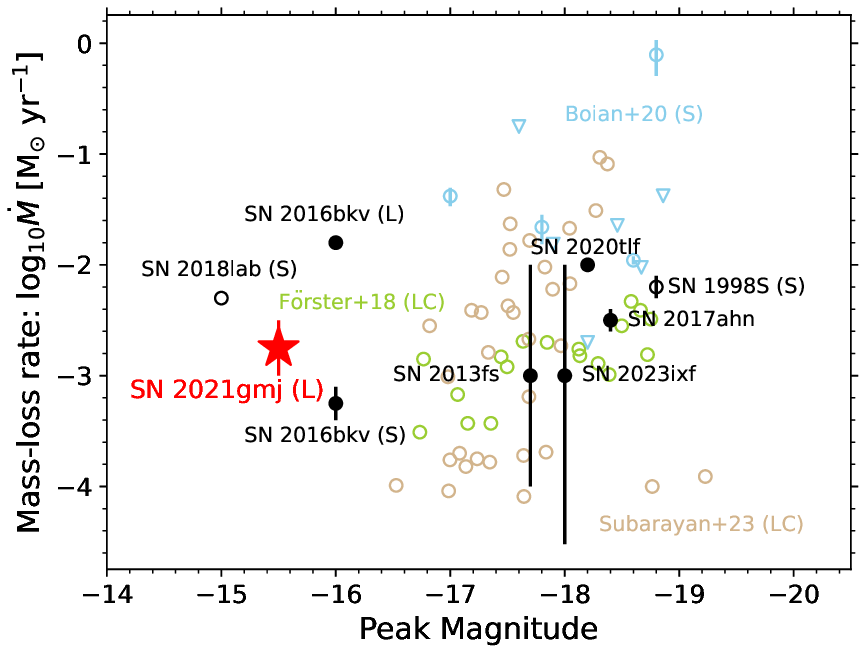}
\caption{
  \label{fig:overview}
The estimated mass-loss rate and peak absolute magnitude relation for 
 SN 2021gmj and other Type II SNe. 
 The sample here includes previous statistical studies (\citealt{Forster2018, Boian2020, Subrayan2023}) 
 and individual studies: SN 1998S (\citealt{Shivvers2015}), SN 2013fs (\citealt{Yaron2017, Moriya2017}), 
 SN 2016bkv (\citealt{nakaoka2018, Deckers2021}), SN 2017ahn (\citealt{Tartaglia2021}), 
 SN 2018lab (\citealt{Pearson2023}), SN 2020tlf (\citealt{Jacobson-Galan2022}), and 
 SN 2023ixf (\citealt{Bostroem2023, Grefenstette2023, Teja2023}). 
For the individual objects, peak absolute magnitude is taken from the Open Supernova Catalog. 
 For the results by \citet{Subrayan2023}, we show only the the objects with a good light curve coverage and a good model fit.
 The bands of peak absolute magnitude are following: 
 $g$ \citep{Forster2018,Subrayan2023}, 
 $r$ and $R$ (\citealt{Boian2020}), 
 $V$ (SN 1998S, 2013fs, 2016bkv, 2017ahn, 2018lab, 2020tlf, 2021gmj, and 2023ixf).
 Mass-loss rate is estimated by analyzing either light curves or spectra. 
 (L) and (S) in the figures represent the methods to estimate of mass-loss rate: (L) is light curves, (S) is spectra. 
 Open circles (triangles) are the estimated mass loss rate (upper limit) using only light curves or spectra
 while filled circles show the estimated mass loss rate using both light curves and spectra. 
}
\end{center}
\end{figure}

Here we compare the estimated mass-loss rates among Type II SNe and 
discuss the progenitor properties.
Figure \ref{fig:overview} shows the estimated mass-loss rate as a function of the peak absolute magnitude.
Our result for SN 2021gmj (red) is compared with the results from other studies including large samples of Type II SNe.
\cite{Forster2018} constrained progenitor mass-loss rates from $10^{-4}$ to 
$10^{-2} ~ \rm \Msun ~ yr^{-1}$ using early part of light curves and multi-wavelength radiative transfer models. 
\cite{Subrayan2023} also estimated mass loss rate for ZTF Type II SNe to be $10^{-4} -$ 
$10^{-1} ~ \rm \Msun ~ yr^{-1}$ using multi-wavelength radiative transfer models.
\cite{Boian2020} analyzed early-time spectra by comparing with the detailed spectral models. The estimated mass-loss rates 
vary from a few $10^{-4}$ to 1 $\rm \Msun ~ yr^{-1}$. 
The results from the other individual studies are also plotted in Figure \ref{fig:overview}. 
Such individual studies mainly focus on normal-luminosity SNe.
The estimated mass-loss rates are $\sim 10^{-3} - 10^{-2} ~ \rm \Msun ~ yr^{-1}$. 

The number of studies for intermediate-luminosity or sub-luminous SNe is still small. 
The mass-loss rate for SN 2016bkv ($M_{V} \sim -16$) is estimated to be 
$\sim 6\times 10^{-4} - 1.7\times 10^{-2} ~ \rm \Msun ~ yr^{-1}$ 
by the analysis of early-time light curves and spectra (\citealt{nakaoka2018, Deckers2021}). 
For SN 2018lab ($M_{V} \sim -15.1$), which is also one of the sub-luminous SNe,
\cite{Pearson2023} estimated the mass-loss rate to be 
$\sim 5\times 10^{-3} ~ \rm \Msun ~ yr^{-1}$ by analyzing early-time spectra. 
Our results on SN 2021gmj newly add an estimate of mass-loss rate in the lower-luminosity SN samples:
$10^{-3} - 10^{-2.5} ~ \rm \Msun ~ yr^{-1}$ (Section \ref{subsec:properties}).
All of these studies so far suggest that, 
as in normal SNe,
low-luminosity SNe also have higher mass-loss rates than typical mass-loss rate of RSGs.
We infer that the progenitor of SN 2021gmj is a relatively low-mass RSG (Section \ref{subsec:M_prog}). 
These results suggest a relatively low-mass SN progenitor can also experience the enhanced mass loss 
just before explosion, which gives a hint to understand the mechanism of the enhanced mass loss. 

%%%%%%%%%%%%%%%%%%%%%%%%%%%%%%%%%%%%%%%%%%%%%%%%%%%% 
% Figure: Compare with other Type II SNe
%%%%%%%%%%%%%%%%%%%%%%%%%%%%%%%%%%%%%%%%%%%%%%%%%%%% 
%%%%%%%%%%%%%%%%%%%%%%%%%%%%%%%%%%%%%%%%%%%%%%%%%%%%

%%%%%%%%%%%%%%%%%%%%%%%%%%%%%%%%%%%%%%%%%%%%%%%%%%%% 
% Section: Conclusions
%%%%%%%%%%%%%%%%%%%%%%%%%%%%%%%%%%%%%%%%%%%%%%%%%%%% 

\section{Conclusions}
\label{sec:conclusions}

In this paper, we presented detailed photometric, 
spectroscopic and imaging polarimetric observations of the intermediate-luminosity Type IIP SN 2021gmj. 
The data were taken from a few days after the explosion to more than a year. 
The photometric observations of SN 2021gmj show that this SN lies at the luminosity range 
between sub-luminous and normal Type IIP SNe. 
The peak absolute $V$-band magnitude of SN 2021gmj is $-15.5$ mag, similar to those of intermediate-luminosity SNe 2009ib, 2009N and 2012A. 
The plateau length of SN 2021gmj is also similar to those of SN 2009N and 2012A. 

The spectral evolution also indicates that SN 2021gmj is an intermediate-luminosity Type IIP SN. 
The spectra of SN 2021gmj have narrow line profiles that are seen in those of sub-luminous SNe and intermediate-luminosity SNe. 
We measure the velocities of H$\alpha$ and H$\beta$ lines and compared them with those of other Type IIP SNe. 
The ejecta velocity of SN 2021gmj is also between normal SNe and sub-luminous SNe. 

Using the tail bolometric luminosity, the \Nifs ~mass of SN 2021gmj is estimated to be about 0.022 $\rm \Msun$. 
This is also intermediate between the normal SNe and sub-luminous SN 2005cs, 
and close to that of SN 2009N. 
Comparison of the nebular emission line ratio of [O I] $\lambda$6300,6364/[Ca II] $\lambda$7291,7324 
with theoretical models (\citealt{Jerkstrand2014}) suggests that the ZAMS mass of SN 2021gmj is about 12 $\rm \Msun$ which is a lower end of massive stars. 

We estimate the properties of the explosion and the CSM by comparing the observed light curves with the light curve models from \cite{Moriya2023}. 
We find that the explosion energy of $3\times 10^{50} ~ \rm erg$ gives a best fit 
to the plateau luminosity. 
The models with the mass-loss rate of $10^{-3}$ to $10^{-2.5} ~ \rm \Msun ~ yr^{-1}$ can reproduce the fast rise of the light curve, 
suggesting that the dense CSM exists around the progenitor of SN 2021gmj before the explosion. 
In the spectra just after the explosion, a broad ledge feature around 4,600 \AA ~is observed. 
This feature is likely caused by a blend of several highly-ionized lines from the CSM. 
It further supports the presence of the CSM around the progenitor of SN 2021gmj. 
From these properties of SN 2021gmj, we suggest that a relatively low-mass SN progenitor of intermediate-luminosity SNe can also experiences 
the enhanced mass loss just before the explosion, as suggested for normal Type IIP SNe. 

%%%%%%%%%%%%%%%%%%%%%%%%%%%%%%%%%%%%%%%%%%%%%%%%%%%% 
% Section: Conclusions
%%%%%%%%%%%%%%%%%%%%%%%%%%%%%%%%%%%%%%%%%%%%%%%%%%%% 

\section*{ACKNOWLEDGEMENTS}
\label{sec:acknowledgements}
We extend our thanks to Anders Jerkstrand for providing theoretical models.
The data from the Seimei and Kanata telescopes were taken under the KASTOR (Kanata And Seimei Transient Observation Regime) project (Seimei program IDs: 21A-N-CT02, 21A-K-0001, 21A-K-0002, 21A-O-0004, 21B-N-CT09, 21B-K-0004, 22A-N-CT09, 22A-K-004). The Seimei telescope at the Okayama Observatory is jointly operated by Kyoto University and the Astronomical Observatory of Japan (NAOJ), with assistance provided by the Optical and Near-Infrared Astronomy Inter-University Cooperation Program. The authors thank the TriCCS developer team (which has been supported by the JSPS KAKENHI grant Nos. JP18H05223, JP20H00174, and JP20H04736, and by NAOJ Joint Development Research).
This work is partly based on observations made under program IDs P63-016 and P65-005 with the Nordic Optical Telescope, owned in collaboration by the University of Turku and Aarhus University, and operated jointly by Aarhus University, the University of Turku and the University of Oslo, representing Denmark, Finland and Norway, the University of Iceland and Stockholm University at the Observatorio del Roque de los Muchachos, La Palma, Spain, of the Instituto de Astrofisica de Canarias. 
We thank the staff of IAO, Hanle, and CREST, Hosakote, that made HCT observations possible. The facilities at IAO and CREST are operated by the Indian Institute of Astrophysics, Bangalore. DKS acknowledges the support provided by DST-JSPS under grant number DST/INT/JSPS/P 363/2022.
This research is partially supported by the Optical and Infrared Synergetic Telescopes for Education and Research (OISTER) program funded by the MEXT of Japan. The authors acknowledge the support by JSPS KAKENHI grant JP21H04997, JP23H00127, JP23H04894 (M.T.), JP20H00174 (K.M.), and by the JSPS Open Partnership Bilateral Joint Research Project between Japan and Finland JPJSBP120229923 (K.M.).

\section*{Data Availability}
The data used in this research will be shared on reasonable request to the corresponding author.

%%%%%%%%%%%%%%%%%%%%%%%%%%%%%%%%%%%%%%%%%%%%%%%%%%%% 
% Reference
%%%%%%%%%%%%%%%%%%%%%%%%%%%%%%%%%%%%%%%%%%%%%%%%%%%% 
%\bibliographystyle{mnras}
%\bibliography{reference}

\vspace{1cm}
\noindent
{\it
$^{1}$Astronomical Institute, Tohoku University, Sendai 980-8578, Japan\\
$^{2}$Division for the Establishment of Frontier Sciences, Organization for Advanced Studies, Tohoku University, Sendai 980-8577, Japan\\
$^{3}$Nishi-Harima Astronomical Observatory, Center for Astronomy, University of Hyogo, 407-2 Nishigaichi, Sayo-cho, Sayo, Hyogo 679-5313, Japan\\
$^{4}$Department of Astronomy, Kyoto University, Kitashirakawa-Oiwake-cho, Sakyo-ku, Kyoto 606-8502, Japan\\
$^{5}$Indian Institute of Astrophysics, Koramangala 2nd Block, Bangalore 560034, India\\
$^{6}$Pondicherry University, Chinna Kalapet, Kalapet, Puducherry 605014, India\\
$^{7}$Hiroshima Astrophysical Science Centre, Hiroshima University, 1-3-1 Kagamiyama, Higashi-Hiroshima, Hiroshima 739-8526, Japan\\
$^{8}$Department of Physics, Graduate School of Advanced Science and Engineering, Hiroshima University, 1-3-1 Kagamiyama, Higashi-Hiroshima, Hiroshima 739-8526, Japan\\
$^{9}$Department of Physics and Astronomy, University of Turku, FI-20014 Turku, Finland\\
$^{10}$Aalto University Mets\"{a}hovi Radio Observatory, Mets\"{a}hovintie 114, 02540 Kylm\"{a}l\"{a}, Finland\\
$^{11}$Aalto University Department of Electronics and Nanoengineering, P.O. BOX 15500, FI-00076 AALTO Finland\\
$^{12}$National Astronomical Observatory of Japan, National Institutes of Natural Sciences, 2-21-1 Osawa, Mitaka, Tokyo 181-8588, Japan\\
$^{13}$School of Physics and Astronomy, Faculty of Science, Monash University, Clayton, Victoria 3800, Australia\\
$^{14}$Astronomical Science Program, Graduate Institute for Advanced Studies, SOKENDAI, 2-21-1 Osawa, Mitaka, Tokyo 181-8588, Japan\\
$^{15}$Department of Physics, Faculty of Science and Engineering, Konan University, 8-9-1 Okamoto, Kobe, Hyogo 658-8501, Japan\\
$^{16}$Kavli Institute for the Physics and Mathematics of the Universe (WPI), The University of Tokyo, 5-1-5 Kashiwanoha, Kashiwa, Chiba 277-8583, Japan\\
$^{17}$Planetary Exploration Research Center, Chiba Institute of Technology, 2-17-1 Tsudanuma, Narashino, Chiba 275-0016, Japan\\
$^{18}$Institute of Astronomy, Graduate School of Science, The University of Tokyo, 2-21-1 Osawa, Mitaka, Tokyo 181-0015, Japan\\
$^{19}$Okayama Observatory, Kyoto University, 3037-5 Honjo, Kamogatacho, Asakuchi, Okayama 719-0232, Japan\\
$^{20}$Department of Multi-Disciplinary Sciences, Graduate School of Arts and Sciences, The University of Tokyo, 3-8-1 Komaba, Meguro, Tokyo 153-8902, Japan\\
$^{21}$Kiso Observatory, Institute of Astronomy, Graduate School of Science, The University of Tokyo, 10762-30 Mitake, Kiso-machi, Kiso-gun, Nagano 397-0101, Japan\\
$^{22}$Okayama Branch Office, Subaru Telescope, NAOJ, NINS, Kamogata, Asakuchi, Okayama 719-0232, Japan\\
$^{23}$Japan Spaceguard Association, Bisei Spaceguard Center, 1716-3 Okura, Bisei, Ibara, Okayama 714-1411, Japan\\
$^{24}$UTokyo Organization for Planetary Space Science, The University of Tokyo, 7-3-1 Hongo, Bunkyo-ku, Tokyo 113-0033, Japan\\
$^{25}$Collaborative Research Organization for Space Science and Technology, The University of Tokyo, 7-3-1 Hongo, Bunkyo-ku, Tokyo 113-0033, Japan\\
$^{26}$Amanogawa Galaxy Astronomy Research Center (AGARC), Graduate School of Science and Engineering, Kagoshima University, 1-21-35 Korimoto, Kagoshima, Kagoshima 890-0065, Japan
}

%%%%%%%%%%%%%%%%%%%%%%%%%%%%%%%%%%%%%%%%%%%%%%%%%%%% 
% Appendix
%%%%%%%%%%%%%%%%%%%%%%%%%%%%%%%%%%%%%%%%%%%%%%%%%%%% 
\appendix
\section{Data tables}
%%%%%%%%%%%%%%%%%%%%%%%%%%%%%%%%%%%%%%%%%%%%%%%%%%%% 
% Table : Optical photometry
%%%%%%%%%%%%%%%%%%%%%%%%%%%%%%%%%%%%%%%%%%%%%%%%%%%% 
\begin{table*}
  \begin{center}
  \caption{Optical photometry of SN 2021gmj}
  \label{tab:opt-photometry}
  \renewcommand{\arraystretch}{0.95}
    \begin{threeparttable}
      \begin{adjustwidth}{-1in}{-1in} \begin{center} \resizebox{1.1\textwidth}{!}{
        \begin{tabular}{cccccccccccc}
        \hline
        \hline
        UT Date & MJD & Phase\tnote{a} & $U$ & $B$ & $g$ & Tomo-e & $V$ & $R$ & $r$ & $i$ & $I$ \\
        (YYYY-MM-DD) & & (days) & (mag) & (mag) & (mag) & (mag) & (mag) & (mag) & (mag) & (mag) & (mag) \\
        \hline
        2021-03-21 & 59294.57 & 1.25 & - & 15.98 (0.029) & - & - & 16.05 (0.037) & 15.79 (0.026) & - & - & 15.66 (0.021) \\
        2021-03-21 & 59294.75 & 1.43 & - & - & - & - & 16.28 (0.11) & 16.13 (0.031) & - & - & 15.98 (0.082) \\
        2021-03-22 & 59295.48 & 2.16 & - & - & - & 15.79 (0.064) & - & - & - & - & - \\
        2021-03-22 & 59295.50 & 2.18 & - & 15.79 (0.054) & - & - & 16.26 (0.046) & 16.11	(0.15) & - & - & 15.95 (0.051) \\
        2021-03-22 & 59295.60 & 2.28 & - & 15.97 (0.023) & - & - & 15.98 (0.020) & 15.71 (0.021) & - & - & 15.56 (0.022) \\
        2021-03-23 & 59296.46 & 3.14 & - & 15.98 (0.027) & - & - & 15.96 (0.022) & 15.68 (0.02) & - & - & 15.52 (0.021) \\
        2021-03-23 & 59296.54 & 3.22 & - & - & - & 15.60 (0.051) & - & - & - & - & - \\
        2021-03-23 & 59296.54 & 3.22 & - & - & - & - & - & 15.89 (0.061) & - & - & 15.97	(0.11) \\
        2021-03-25 & 59298.52 & 5.21 & - & - & - & - & 16.12	(0.074) & 15.83 (0.044) & - & - & 15.72 (0.019) \\
        2021-03-25 & 59298.68 & 5.37 & - & 15.95 (0.022) & - & - & 15.89 (0.023) & 15.56 (0.023) & - & - & 15.38 (0.020) \\
        2021-03-26 & 59299.43 & 6.11 & - & - & - & 15.72 (0.083) & - & - & - & - & - \\
        2021-03-26 & 59299.50 & 6.18 & 15.17 (0.069) & 15.79 (0.048) & - & - & 15.80 (0.034) & 15.75 (0.008) & - & - & 15.54 (0.020) \\
        2021-03-26 & 59299.60 & 6.28 & - & 16.03	(0.025) & - & - & 15.93 (0.021) & 15.56 (0.027) & - & - & 15.38 (0.025) \\
        2021-03-29 & 59302.49 & 9.17 & - & - & - & 15.65 (0.041) & - & - & - & - & - \\
        2021-03-29 & 59302.50 & 9.18 & - & 15.88 (0.075) & - & - & 15.85 (0.103) & 15.58 (0.021) & - & - & 15.54 (0.022) \\
        2021-03-29 & 59302.67 & 9.35 & - & 16.04	(0.027) & - & - & 15.96 (0.022) & 15.50 (0.023) & - & - & 15.31 (0.024) \\
        2021-03-30 & 59303.45 & 10.13 & - & 16.05 (0.021) & - & - & 15.97 (0.026) & 15.53 (0.021) & - & - & 15.36 (0.022) \\
        2021-03-30 & 59303.50 & 10.18 & - & 15.90 (0.065) & - & - & 15.88 (0.022) & 15.56 (0.032) & - & - & 15.52 (0.040) \\
        2021-04-01 & 59305.48 & 12.16 & - & 16.11 (0.022) & - & - & 15.98 (0.020) & 15.49 (0.021) & - & - & 15.31	 (0.021) \\
        2021-04-01 & 59305.50 & 12.18 & - & 16.00 (0.015) & - & - & 15.88 (0.007) & 15.55 (0.010) & - & - & 15.48 (0.031) \\
        2021-04-02 & 59306.52 & 13.20 & - & 16.13 (0.021) & - & - & 15.98 (0.020) & 15.48 (0.020) & - & - & 15.28 (0.020) \\
        2021-04-05 & 59309.50 & 16.18 & 15.81 (0.199) & 16.08 (0.025) & - & - & 15.88 (0.011) & 15.49 (0.012) & - & - & 15.39 (0.017) \\
        2021-04-05 & 59309.57 & 16.25 & - & - & - & 15.75 (0.002) & - & - & - & - & - \\
        2021-04-05 & 59309.66 & 16.34 & - & 16.19 (0.021) & - & - & 15.96 (0.021) & 15.45 (0.020) & - & - & 15.23 (0.020) \\
        2021-04-06 & 59310.74 & 17.42 & - & 16.22 (0.020) & - & - & 15.99 (0.020) & 15.45 (0.020) & - & - & 15.26 (0.021) \\
        2021-04-07 & 59311.72 & 18.40 & - & 16.28 (0.025) & - & - & 15.97 (0.026) & 15.46 (0.022) & - & - & 15.26 (0.021) \\
        2021-04-08 & 59312.76 & 19.44 & - & 16.36 (0.040) & - & - & 16.02 (0.023) & 15.48 (0.020) & - & - & 15.26 (0.020) \\
        2021-04-09 & 59313.78 & 20.46 & - & 16.34 (0.023) & - & - & 16.01 (0.020) & 15.46 (0.020) & - & - & 15.24 (0.020) \\
        2021-04-10 & 59314.51 & 21.20 & - & - & - & 15.93 (0.042) & - & - & - & - & - \\
        2021-04-10 & 59314.69 & 21.37 & - & 16.37 (0.021) & - & - & 16.02 (0.020) & 15.48 (0.020) & - & - & 15.25 (0.022) \\
        2021-04-11 & 59315.50 & 22.18 & 16.39 (0.160) & 16.37 (0.021) & - & - & 15.93 (0.012) & 15.54 (0.013) & - & - & 15.42 (0.048) \\
        2021-04-11 & 59315.55 & 22.23 & - & 16.41 (0.022) & - & - & 16.04 (0.021) & 15.49 (0.020) & - & - & 15.27 (0.021) \\
        2021-04-14 & 59318.45 & 25.13 & - & - & - & 15.94 (0.081) & - & - & - & - & - \\
        2021-04-15 & 59319.56 & 26.24 & - & 16.52 (0.020) & - & - & 16.06 (0.020) & 15.51 (0.021) & - & - & 15.25 (0.020) \\
        2021-04-18 & 59322.70 & 29.38 & - & 16.52 (0.033) & - & - & 16.07 (0.025) & 15.52 (0.022) & - & - & 15.23 (0.020) \\
        2021-04-19 & 59323.68 & 30.36 & - & 16.58 (0.026) & - & - & 16.07 (0.022) & 15.52 (0.021) & - & - & 15.24 (0.021) \\
        2021-04-20 & 59324.50 & 31.18 & - & - & - & 15.86 (0.097) & - & - & - & - & - \\
        2021-04-20 & 59324.58 & 31.26 & - & 16.62 (0.020) & - & - & 16.08 (0.021) & 15.52 (0.021) & - & - & 15.23 (0.020) \\
        2021-04-23 & 59327.50 & 34.18 & - & 16.62 (0.097) & - & - & 15.97 (0.020) & 15.55 (0.022) & - & - & 15.33 (0.035) \\
        2021-04-26 & 59330.50 & 37.18 & - & 16.69 (0.029) & - & - & 15.95 (0.022) & 15.55 (0.005) & - & - & - \\
        2021-04-27 & 59331.57 & 38.25 & - & - & - & 15.88 (0.041) & - & - & - & - & - \\
        2021-05-09 & 59343.53 & 50.21 & - & - & - & 15.91 (0.051) & - & - & - & - & - \\
        2021-05-13 & 59347.58 & 54.26 & - & - & - & 15.90 (0.012) & - & - & - & - & - \\
        2021-05-14 & 59348.53 & 55.21 & - & - & - & 16.17 (0.032) & - & - & - & - & - \\
        2021-05-15 & 59349.54 & 56.22 & - & - & - & 15.77 (0.14) & - & - & - & - & - \\
        2021-05-22 & 59356.57 & 63.25 & - & - & - & 16.19 (0.14) & - & - & - & - & - \\
        2021-05-22 & 59356.58 & 63.26 & - & - & - & - & 16.04 (0.045) & 15.69 (0.025) & - & - & 15.36 (0.070) \\
        2021-05-23 & 59357.49 & 64.17 & - & - & - & 15.88 (0.026) & - & - & - & - & - \\
        2021-05-25 & 59359.50 & 66.18 & - & 17.03 (0.098) & - & - & 15.97 (0.032) & 15.53 (0.033) & - & - & 15.24 (0.038) \\
        2021-05-29 & 59363.63 & 70.31 & - & 16.98 (0.021) & - & - & 16.1 (0.020) & 15.48 (0.021) & - & - & 15.09 (0.023) \\
        2021-05-30 & 59364.48 & 71.16 & - & - & - & 16.04 (0.096) & - & - & - & - & - \\
        2021-06-01 & 59366.47 & 73.15 & - & - & - & 15.87 (0.056) & - & - & - & - & - \\
        2021-06-09 & 59374.48 & 81.16 & - & - & - & - & 16.26 (0.011) & - & - & - & 15.35 (0.021) \\
        2021-06-10 & 59375.50 & 82.18 & - & - & - & - & 16.04 (0.180) & - & - & - & - \\
        2021-06-13 & 59378.50 & 85.18 & - & 17.37 (0.018) & - & - & 16.13 (0.015) & 15.65 (0.016) & - & - & 15.34 (0.023) \\
        2021-06-18 & 59383.50 & 90.18 & - & 17.54 (0.074) & - & - & 16.22 (0.029) & 15.68 (0.014) & - & - & 15.40 (0.015) \\
        2021-06-19 & 59384.52 & 91.20 & - & - & - & - & - & 15.78	(0.038) & - & - & - \\
        2021-06-20 & 59385.55 & 92.23 & - & 17.29 (0.022) & - & - & 16.34 (0.021) & 15.68 (0.020) & - & - & 15.28 (0.027) \\
        2021-06-22 & 59387.50 & 94.18 & - & 17.55 (0.117) & - & - & 16.30 (0.018) & 15.76 (0.018) & - & - & 15.44 (0.060) \\
        2021-06-29 & 59394.50 & 101.18 & - & 17.93 (0.020) & - & - & 16.48 (0.022) & 15.89 (0.020) & - & - & 15.60 (0.068) \\
        2021-07-12 & 59407.54 & 114.22 & - & 18.38 (0.066) & - & - & 18.29	(0.049) & 17.10 (0.023) & - & - & 16.55 (0.025) \\
        2021-07-14 & 59409.50 & 116.18 & - & 19.48 (0.170) & - & - & - & 17.31 (0.016) & - & - & 16.87 (0.034) \\
        2021-07-19 & 59414.51 & 121.19 & - & - & - & - & 18.32 (0.020) & - & - & - & 16.78 (0.020) \\
        2021-07-28 & 59423.49 & 130.17 & - & - & - & - & - & - & - & - & 16.96 (0.023) \\
        2021-08-05 & 59431.46 & 138.15 & - & 18.53 (0.13) & - & - & 18.47 (0.037) & 17.57 (0.21) & - & - & 16.96 (0.027) \\
        2021-09-29 & 59486.82 & 193.50 & - & - & - & - & - & - & 18.52 (0.078) & - & - \\
        2021-10-10 & 59497.50 & 204.18 & - & - & - & - & - & 18.30 (0.056) & - & - & -\\
        2021-10-21 & 59508.50 & 215.18 & - & - & - & - & 19.65 (0.102) & 18.40 (0.051) & - & - & 18.13 (0.054) \\
        2021-10-25 & 59512.80 & 219.48 & - & - & 19.94 (0.14) & - & - & - & 18.73 (0.072) & - & - \\
        2021-11-08 & 59526.50 & 233.18 & - & - & - & - & 19.58 (0.045) & 18.46 (0.029) & - & - & 18.19 (0.047) \\
        2021-11-20 & 59538.50 & 245.18 & - & - & - & - & 19.59 (0.077) & 18.47 (0.034) & - & - & 18.32 (0.045) \\
        2021-12-02 & 59550.85 & 257.53 & - & - & - & - & - & - & - & 18.28 (0.079) & - \\
        2021-02-10 & 59620.63 & 327.31 & - & - & - & - & - & - & 19.13 (0.25) & - & - \\
        2022-03-06 & 59644.50 & 351.18 & - & - & - & - & - & - & 19.34 (0.023) & 19.18 (0.056) & - \\
        \hline
        \end{tabular}
        } \end{center} \end{adjustwidth}
      \begin{tablenotes}
      \item[a] Days after the discovery date (MJD 59293.319)
      \end{tablenotes}
    \end{threeparttable}
  \end{center}
\end{table*}
%%%%%%%%%%%%%%%%%%%%%%%%%%%%%%%%%%%%%%%%%%%%%%%%%%%% 
%%%%%%%%%%%%%%%%%%%%%%%%%%%%%%%%%%%%%%%%%%%%%%%%%%%% 
% Table : NIR photometry
%%%%%%%%%%%%%%%%%%%%%%%%%%%%%%%%%%%%%%%%%%%%%%%%%%%% 
\begin{table*}
  \begin{center}
  \caption{NIR photometry of SN 2021gmj}
  \label{tab:nir-photometry}
    \begin{threeparttable}
      \begin{tabular}{cccccc}
      \hline
      \hline
      UT Date & MJD & Phase & $J$ & $H$ & $K_{s}$ \\
      (YYYY-MM-DD) & & (days) & (mag) & (mag) & (mag) \\
      \hline
      2021-03-21 & 59294.75 & 1.43 & 16.10 (0.098) & - & - \\
      2021-03-22 & 59295.50 & 2.18 & 15.94 (0.11) & 15.70 (0.077) & - \\
      2021-03-23 & 59296.54 & 3.22 & 15.76 (0.077) & 15.62 (0.088) & 15.47 (0.21) \\
      2021-03-25 & 59298.52 & 5.21 & 15.44 (0.065) & 15.50 (0.038) & 15.12 (0.17) \\
      2021-05-22 & 59356.58 & 63.26 & 14.84 (0.049) & 14.58 (0.046) & 14.59 (0.076) \\
      2021-06-09 & 59374.48 & 81.16 & 14.89 (0.031) & 14.62 (0.066) & 14.65 (0.14) \\
      2021-06-19 & 59384.52 & 91.20 & 15.19 (0.010) & - & - \\
      2021-06-22 & 59387.52 & 94.20 & 15.31 (0.042) & 14.88 (0.054) & - \\
      \hline
      \end{tabular}
      \begin{tablenotes}
      \item[a] Days after the discovery date (MJD 59293.319)
      \end{tablenotes}
    \end{threeparttable}
  \end{center}
\end{table*}
%%%%%%%%%%%%%%%%%%%%%%%%%%%%%%%%%%%%%%%%%%%%%%%%%%%% 
%%%%%%%%%%%%%%%%%%%%%%%%%%%%%%%%%%%%%%%%%%%%%%%%%%%% 
% Table : Spectroscopic obs
%%%%%%%%%%%%%%%%%%%%%%%%%%%%%%%%%%%%%%%%%%%%%%%%%%%% 
\begin{table*}
  \begin{center}
  \caption{Log of spectroscopic observations of SN 2021gmj}
  \label{tab:spectra}
  \begin{threeparttable}
    \begin{tabular}{cccccc}
    \hline
    \hline
    UT Date & MJD & Phase\tnote{a} & Instrument & Coverage & Resolution \\
    (YYYY-MM-DD) & & (days) & & ({\AA}) & \\
    \hline
    2021-03-21 & 59294.9 & 2.6 & KOOLS-IFU & 4100-8900 & 500 \\
    2021-03-24 & 59297.0 & 4.7 & KOOLS-IFU & 4100-8900 & 500 \\
    2021-03-26 & 59299.9 & 7.6 & HFOSC & 3500-7800 & 800 \\
    2021-03-29 & 59302.7 & 10.4 & HFOSC & 3500-9000 & 800 \\
    2021-03-30 & 59303.7 & 11.4 & HFOSC & 3500-9000 & 800 \\
    2021-04-06 & 59310.0 & 17.8 & KOOLS-IFU & 4100-8900 & 500 \\
    2021-04-11 & 59315.8 & 23.5 & HFOSC & 3500-9000 & 800 \\
    2021-04-15 & 59319.0 & 26.7 & KOOLS-IFU & 4100-8900 & 500 \\
    2021-04-23 & 59327.9 & 35.6 & HFOSC & 3500-7800 &800 \\
    2021-04-26 & 59330.7 & 38.4 & HFOSC & 3500-9000 & 800 \\
    2021-05-02 & 59336.8 & 44.5 & KOOLS-IFU & 4100-8900 & 500 \\
    2021-05-14 & 59348.8 & 56.5 & HFOSC & 3500-9000 & 800 \\
    2021-05-25 & 59359.7 & 67.4 & HFOSC & 3500-9000 & 800 \\
    2021-05-29 & 59363.9 & 71.6 & KOOLS-IFU & 4100-8900 & 500 \\
    2021-06-13 & 59378.7 & 86.4 & HFOSC & 3500-9000 & 800 \\
    2021-06-22 & 59387.7 & 95.4 & HFOSC & 3500-9000 & 800 \\
    2022-04-09 & 59678.1 & 385.8 & ALFOSC & 3200-9600 & 360 \\
    \hline
    \end{tabular}
    \begin{tablenotes}
    \item[a] Days after the explosion (MJD 59292.321)
    \end{tablenotes}
  \end{threeparttable}
  \end{center}
\end{table*}
%%%%%%%%%%%%%%%%%%%%%%%%%%%%%%%%%%%%%%%%%%%%%%%%%%%% 
%%%%%%%%%%%%%%%%%%%%%%%%%%%%%%%%%%%%%%%%%%%%%%%%%%%% 
% Table : Imaging ploarimetry
%%%%%%%%%%%%%%%%%%%%%%%%%%%%%%%%%%%%%%%%%%%%%%%%%%%% 
\begin{table*}
  \begin{center}
  \caption{Log of imaging polarimetry of SN 2021gmj}
  \label{tab:polarimetry}
  \begin{threeparttable}
    \begin{tabular}{ccccccc}
    \hline
    \hline
    UT Date & MJD & Phase\tnote{a} & $q_{\rm obs}$ ($V$) & $u_{\rm obs}$ ($V$) & $q_{\rm obs}$ ($R$) & $u_{\rm obs}$ ($R$) \\
    (YYYY-MM-DD) & & (days) & & &  & \\
    \hline
    2021-04-03 & 59307.05 & 14.7 & 0.89$\pm$0.08 & -0.50$\pm$0.14 & 0.76$\pm$0.10 & -0.02$\pm$0.07 \\
    2021-04-27 & 59331.97 & 39.6 & 0.43$\pm$0.11 & -0.40$\pm$0.15 & 1.12$\pm$0.21 & 0.17$\pm$0.17 \\
    2021-05-28 & 59362.97 & 70.6 & -0.02$\pm$0.09 & -0.05$\pm$0.09 & 0.52$\pm$0.13 & -0.08$\pm$0.14 \\
    2021-06-25 & 59390.91 & 98.6 & 0.43$\pm$0.19 & -0.14$\pm$0.34 & -0.11$\pm$0.15 & -0.78$\pm$0.14 \\
    2021-07-01 & 59396.90 & 104.6 & - & - & 0.13$\pm$0.12 & 0.74$\pm$0.20 \\
    2021-07-09 & 59404.89 & 112.6 & - & - & 0.95$\pm$1.06 & 1.77$\pm$0.34  \\
    \hline
    \end{tabular}
    \begin{tablenotes}
    \item[a] Days after the explosion (MJD 59292.321)
    \end{tablenotes}
  \end{threeparttable}
  \end{center}
\end{table*}
%%%%%%%%%%%%%%%%%%%%%%%%%%%%%%%%%%%%%%%%%%%%%%%%%%%% 

% Don't change these lines
\bsp	% typesetting comment
\label{lastpage}
\end{document}